# Polymer Nanocomposites: synthesis and characterization

Anil Arya, A. L. Sharma*

*Department of Physical Sciences, Central University of Punjab, Bathinda-151001, Punjab INDIA*

**Corresponding Author: alsharma@cup.edu.in*

**Abstract**

This chapter deals with the fundamental properties of polymer nanocomposites (PNC) and their characteristics that play a significant role in deciding their capability for the advanced energy storage device. The various synthesization methods used for the preparation of the polymer electrolytes are described followed by the characterization techniques used for the analysis. The properties of the polymer host, salt, nanofiller, ionic liquid, plasticizer and nanoclay/nanorod/nanowire are described. Various ion transport mechanism with different nanoparticle dispersion in polymer electrolytes are highlighted. The various important results are summarized and a pathway is built to fulfill the dream of the future renewable source of energy that is economical and environmental benign. Chapter motivation is focused on the investigation of the role of polymer host, aspect ratio, surface area, nanoparticle shape and size in terms of boosting the electrolytic/electrochemical properties of PNC. It will certainly help in order to open new doors toward the development of the advanced polymeric materials with overall balancing property for enhancement of the fast solid state ionic conductor which would be revolutionized the energy storage/conversion device technology.

## 1. Introduction

In the world, the most of the energy demand (~75 %) of human beings is till now fulfilled by the non-renewable energy resources that include oils, coal, natural gas etc. They are sufficient to provide us energy for a long period but two major drawbacks are associated with them are lack of time and pollution. The combustion of these non-renewable sources of energy has fired the much increase in the $CO_2$ emission (32190 Metric Ton per year) and one day that will cross the global point above which it would become difficult to sustain life on earth. Further, the natural disasters such as storms, floods lead to collapsing of the buildings and public as well government property and it leads to the energy blackout. So, it becomes important to resolve this issue and best alternative is focused. It can be summarized in one line that the environmental change lead by it is affecting our lives and become a necessity to look at some other efficient alternative energy resources. The best appropriate alternative which seems to be feasible is the use of renewable energy such as solar energy, wind energy, hydro energy, electrochemical energy and nuclear energy. In last three decades, SONY commercialized its lithium-ion battery in 1991 and a lot of research is being done till 2016 ($25^{th}$ anniversary of LIB introduction) to replace this non-renewable sources with a renewable one. First time Prof. Michel Armand in 1970 formulated the idea of the intercalation compounds and highlighted the ion migration in between electrodes (rocking chair battery. The battery is now the crucial part of the portable consumer electronics and its demand is supposed to rise in the coming future. Beside this, the lithium-ion-batteries will also be replaced in electric vehicles such as cars, buses, and trains to fight against the pollution. So, the many companies are in frontrunner in boosting the e-mobility and plug-in vehicles (Figure 1 a). This is an environmentally friendly and safe source that can be used for a longer period [1-5].

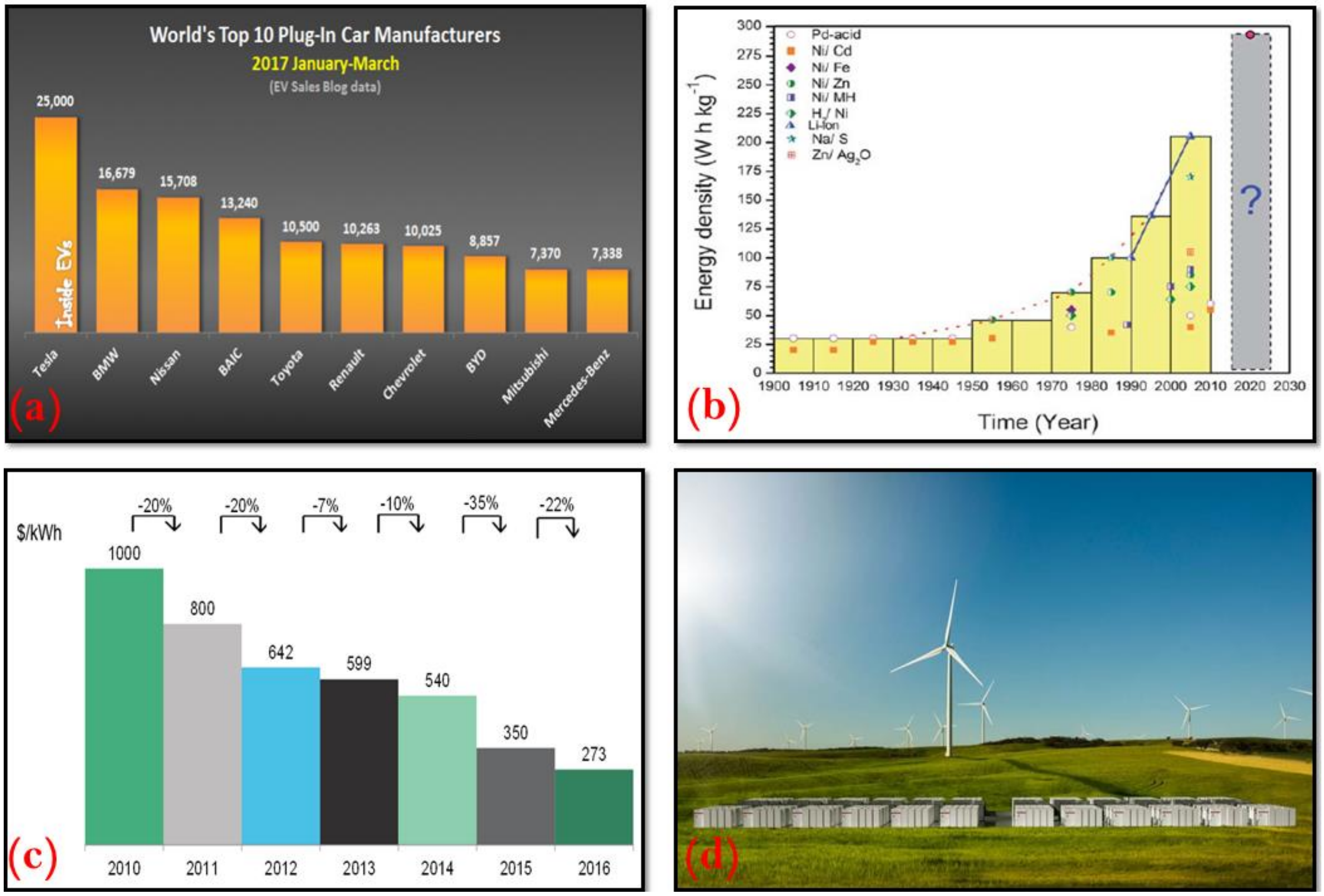

Figure 1. (a) World's Top 10 Plug-In Car Manufacturers – 2017 March (data source: EV Sales Blog) [6], (b) History of development of secondary batteries in view of energy density [7] (c) BNEF lithium-ion battery price survey, 2010-16 ($/kWh) [8] and (d) The world's largest grid-scale battery - the Hornsdale Power Reserve battery [9].

Figure 1 b shows the growth of the commercial secondary batteries from 1950 to 2010 increased by about 3 Wh $kg^{-1}$ per year on average (shown in the dashed line). Dash line shows the progress of last 80 years and the solid line represents the development of Li-ion batteries in last 20 years. This demonstrates that the present energy density (210 $Whkg^{-1}$) will reach up to the target energy densities 500 Wh $kg^{-1}$ and 700 Wh $kg^{-1}$ will be realized in years of 2110 and 2177 respectively. Further another step has been taken to enhance the utilization of the Lithium-ion battery by lowering the prices of the LIB (Figure 1 c).

So, to fulfill the demand of the renewable source of energy, in 2015 first time Tesla disclosed the stationary storage products for homes that boosted the demand for the batteries. The first step toward the renewable energy source is completed in November 2017, Tesla has completed the construction of the world's biggest lithium-ion battery (LIB) with 100 MW capacity just outside the South Australian city of Jamestown for South Australia (Figure 1 d). It will deliver power to the 30,000 homes for about an hour [6-9]. It opened the doors of the future affordable renewable energy source with a lot of possibilities that will revolutionize the energy sector. Another remarkable point with this battery was highlighted in a statement by the Elon Musk, "*You can essentially charge up the battery packs when you have excess power when the cost of production is very low and then discharge it when the cost of power production is high, and this effectively lowers the average cost to the end customer*".

A battery is a very catchy system where, separator/electrolyte is sandwiched between the two different counter electrodes (i) cathode (positive electrode), (ii) anode (negative electrode). The separator/electrolyte provide the path for shuttling of the ion between the electrodes while electrons move via the external circuit. The cathode is generally made of the metal oxide ($LiCoO_2$, $LiCoMnO_2$) and the anode of graphite. During discharge, the ions flow from the anode to the cathode (oxidation at the anode; loss of electrons) through the electrolyte and separator; while during the charging process, charge reverses the direction and the ions flow from the cathode to the anode (reduction of the cathode; a gain of electrons) [10]. Figure 2 demonstrates the operation of a Lithium-ion battery and its components cathode, anode & electrolyte. The role of the electrolyte is to physically separate both electrodes and provides medium for ion migration.

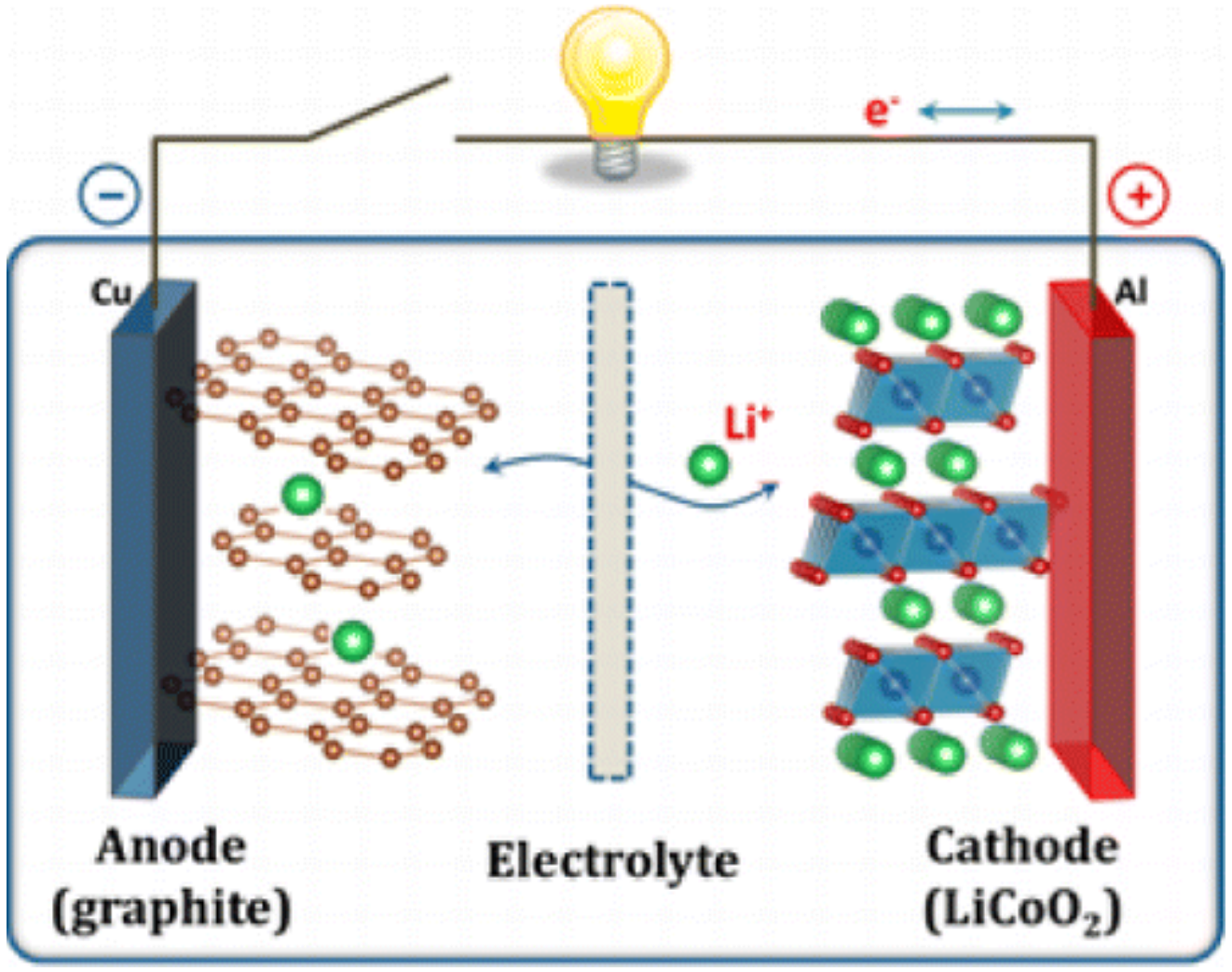

Figure 2. Schematic illustration of the first Li-ion battery (LiCoO2/Li+ electrolyte/graphite). With permission from Ref. [10] Copyright © 2013 American Chemical Society.

A lot of research is going on the development of both cathode and anodes that may provide large energy density and power density without affecting its stability or cycle life. Beside them, electrolyte is a more interesting candidate as it is placed in between the electrodes and it remains always in the active state either it is discharging or charging process. The electrolyte is the heart of the battery and plays a key role in the operation of a battery. Now, a day's most of the battery systems are based on the liquid electrolyte. Although the battery possesses high ionic conductivity but, poor mechanical strength and the stability prevents its use in the commercial applications. Another critical drawback is the dendrite growth formation that leads to short-circuiting of the battery. Another issue is the capacity fading and the narrow safety window due to the liquid electrolyte. So, to overcome all issues faced by the liquid electrolyte based storage system, the most attractive approach which is adopted now days is the use of solid polymer electrolyte (SPE). It prevents the use of separate casing for the electrolyte and it plays a dual role that automatically reduces both cost and weight. Another fundamental advantage with the SPE is that the dendrite growth formation could be minimized at practical level due to good interface contact with the electrodes [11]. SPE is superior in comparison to both liquid polymer electrolyte and the gel polymer electrolyte in many aspects such as stability, flexibility, shape variation, safety, and the cost. Although the SPE is a suitable candidate as an alternative to the conventional electrolyte still some drawback exists there. Another crucial point is the ease of the preparation. It involves the dissolution of host polymer having an electron rich group (polar group) and the salt with bulky anion dissolves in the solvent. The host polymer provides the coordinating sites that favor the fast ion migration and further supported by the segmental motion of the polymer chains. The segmental motion of the polymer chain is linked with the flexibility of the polymer chain, as it pushes the ion from one site to next and mobility is enhanced. The most important one is the low ionic conductivity as compared to the desire for the practical applications and another one is the mechanical stability. So, a new type of solid-state advanced material needs to be a development which can provide us the desirable conductivity value for practical applications (~$10^{-3}$ S $cm^{-1}$) [12-13].

Nowadays, due to increased demand for the Li-ion batteries globally, it becomes important to develop new technologies which can provide the safe and advanced energy storage system. As the nanoparticles (nanofiller, nanorod, nanowire) attractive candidate for developing all components of the battery. One important point to be noticed here that the nanomaterial have possibilities to fulfill the dream of the energy storage system with high energy density and the power density. As it is well known that the shape of the nanoparticle influence strongly all properties. Figure 3 depicts that the next generation energy storage system probably the battery must be of a smaller size with improved performances so that the empty space in the energy storage devices can be fulfilled [14].

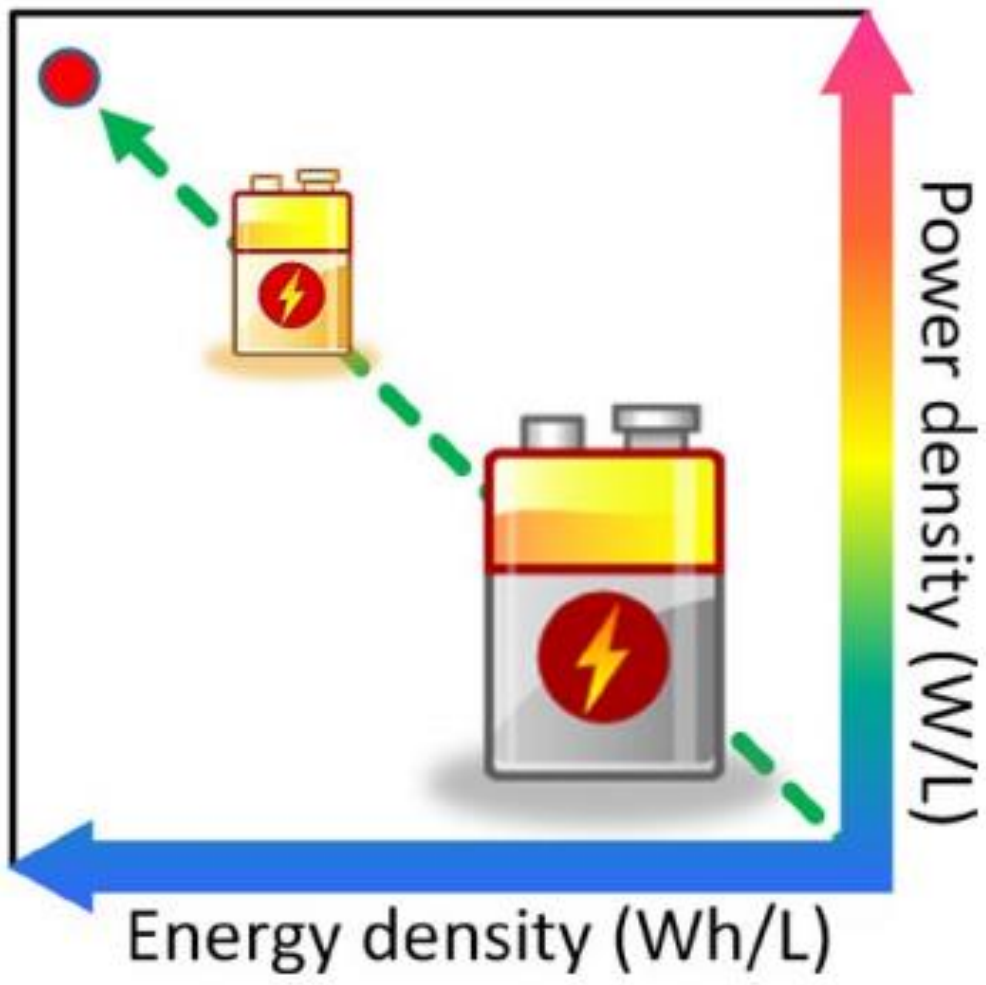

Figure 3. The illustration to a demonstration that future Li-ion batteries should be light and small without any compromise on energy and power. With permission from Ref. [14] Copyright © 2015 John Wiley and Sons.

A number of reviews are published till now with a focus on different types of polymer electrolytes [15-19]. Song et al., highlighted the advantages and characteristics of gel polymer electrolyte for lithium-ion batteries. The chapter covered the four plasticized systems with main focus on, i.e., poly (ethylene oxide) (PEO), poly(acrylonitrile) (PAN), poly(methyl methacrylate) (PMMA), and poly(vinylidene fluoride) (PVdF) based electrolytes [20]. Stephan et al., chapter was focused toward the state-of-art of polymer electrolytes in view of their electrochemical and physical properties for the applications in lithium batteries with main focus on the polymer poly(ethylene oxide) (PEO), poly(acrylonitrile) (PAN), poly(methyl methacrylate) (PMMA), poly(vinylidene fluoride) (PVdF) and poly(vinylidene fluoride-hexafluoro propylene) (PVdF-HFP) as electrolytes. The cycling behavior of $LiMn_2O_4$/polymer electrolyte (PE)/Li cells is also described [21]. Another review by Zhang et al., covered the separators used in liquid electrolyte Li-ion batteries. The classification of separators was done on basis of the structure and composition of the membranes followed by a discussion on the manufacture, characteristics, performance and modifications of the separators [22]. Stephan et al., discussed the composite polymer electrolytes (CPE) with the main focus on electrochemical and physical properties for the applications in lithium batteries. The polymer host discussed were poly (ethylene oxide) (PEO), poly(acrylonitrile) (PAN), poly(methyl methacrylate) (PMMA) and poly(vinylidene fluoride) (PVdF) [23]. The above reviews mostly covered the gel and liquid polymer electrolytes with the main focus on the properties of the polymer electrolytes. To the best of our knowledge, no review articles with the main focus on the shape of the nanofiller in polymer electrolyte have been systematically concluded till now. Polymer nanocomposites are basically the solid polymer electrolytes which are a two-phase system, the first phase acts as host matrix in which different nanoparticles are dispersed [24]. The polymer nanocomposite enables us to develop polymer electrolyte with improved mechanical, thermal properties, voltage stability window and electrochemical properties.

In this Chapter, a brief summary of polymer nanocomposites (PNCs) with nanoparticles of various shape are summarized in detail. First, the status of the LIB technology in the current practical applications is described and followed by the working principle of the battery. Then we discussed the characteristics of the polymer electrolytes and the properties of the constituents used, i.e. polymer host, salt, ionic liquid, plasticizer, nanoclay, nanofiller, nanorod, and nanowire. Then the preparation methods and the common characterization techniques used to identify the suitability of the polymer nanocomposites are discussed followed by the factors influenced by the addition of different nanoparticle. Finally, we have summarized the recent key developments done till now in the field of the solid polymer electrolytes.

## 2. Polymer Electrolytes

In energy storage devices polymer electrolyte plays the dual role of both electrolytes as well separator and is sandwiched between the electrodes. So, the polymer host which is to be used have some special characteristics that make its candidature stronger as compared to other. The polymer electrolytes are classified into three type, (i) *gel polymer electrolytes*: an organic solvent is added in the polymer matrix, (ii) *solid polymer electrolyte*: polymer matrix acts as host matrix and provides coordinating sites for cation migration, no organic solvent required, and (iii) *composite polymer electrolyte*: here nanoparticle is dispersed in the polymer salt matrix, also known as polymer nanocomposites. Figure 4 depicts the properties that are influenced by the addition of the nanoparticle and the synthesization techniques. These properties are linked with one another. The ionic conductivity is linked to the glass transition temperature and hence the flexibility. As ion transport in the polymer, electrolytes is supposed via the amorphous phase so crystallinity associated with the polymer matrix is minimized by altering the polymer chain rearrangement. The cation transference number and ion transference number are crucial for getting insights of applicability of any system for the battery. As during fabrication battery's system is under the stress so, it needs to be focused along with the electrical properties [12, 15, 19]. There are various synthesization approaches for solid polymer electrolytes. The properties of the polymer nanocomposites are also affected by the preparation method.

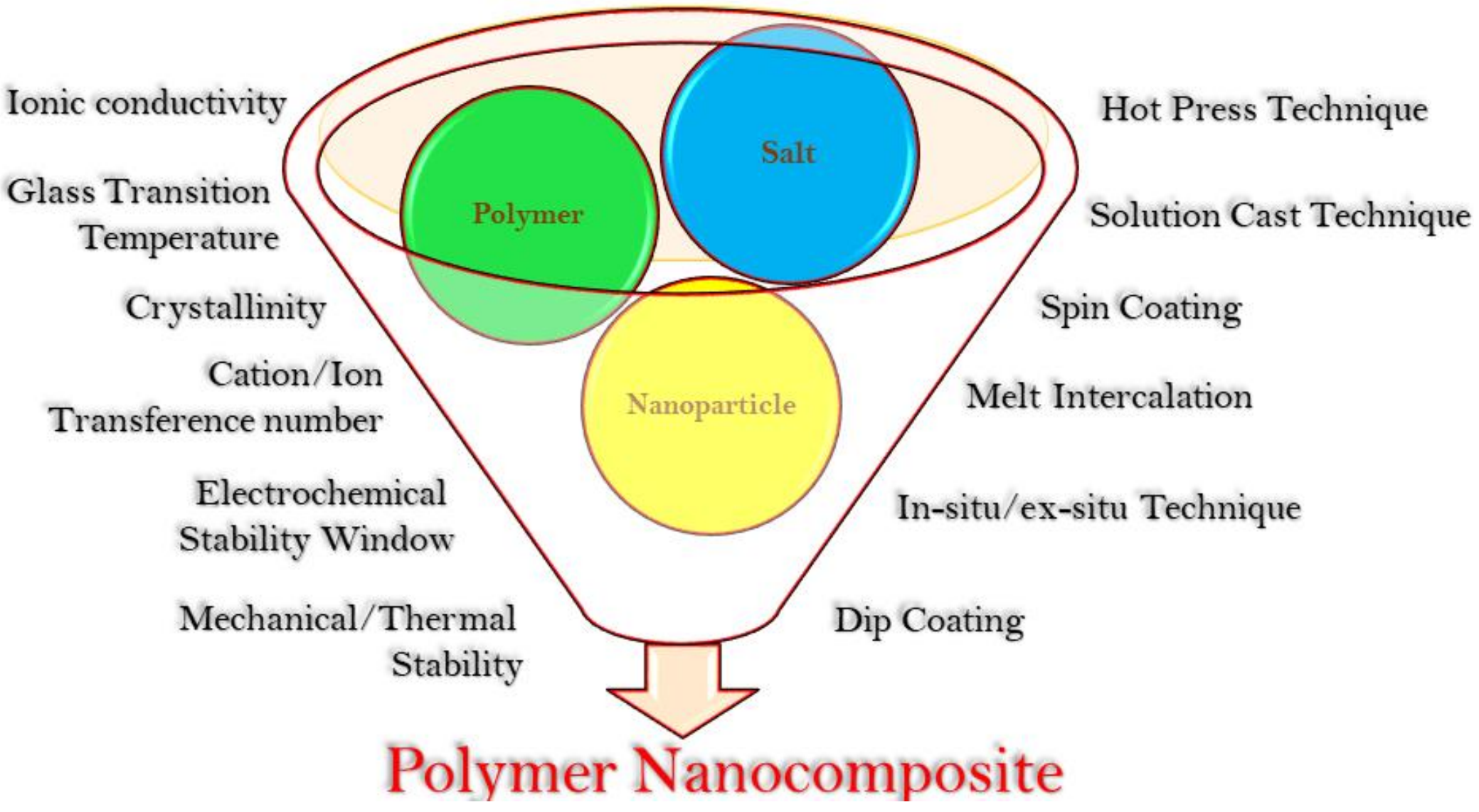

Figure 4. Representation of the properties required for the suitable polymer electrolyte and the synthesization techniques.

As from three decades, a lot of research has been done on the salt/ionic liquid/plasticizer based polymer electrolytes. Although there was a desirable enhancement in the electrical parameters, poor mechanical strength limits their use in commercial applications. So, their alternative was developed and incorporation of nanoparticle was adopted as the most fascinating approach to develop the advanced polymer electrolytes. The suitable nanoparticle adopted were nanofiller, nanoclay and reported enhancement in the electrical as well as thermal/mechanical properties. As nanoparticle shape plays an effective role due to the interconnection of the electrical properties of the nanoparticle shape and surface group. Recently research is focused toward the development of nanoparticle with the high specific surface area. The two approaches were adopted one is the addition of nanorod and another is the addition of nanowire. Both are effective nanoparticle as a long continues continuous path is created that is beneficial for achieving fast ion transport.

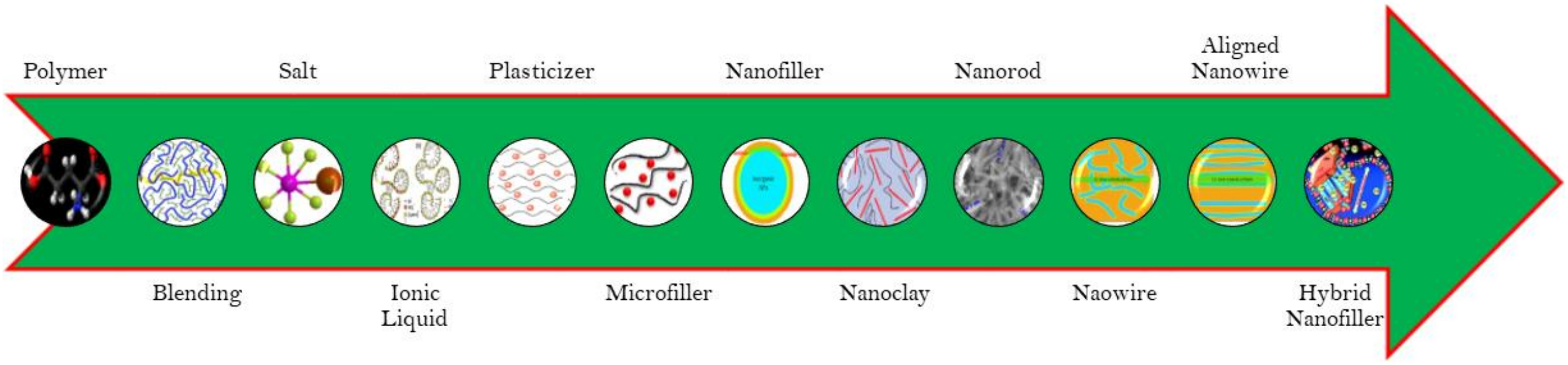

Figure 5. Approaches adopted for the modification of the polymer electrolyte and the progress made till now.

The various approaches used for enhancement of the properties to fulfill the criteria of a solid state ionic conductor are displayed in Figure 5 (From left to right). As the various constituents play a different role in enhancing the electrical properties, thermal properties, and mechanical properties. Table 1 shows the important characteristics that are the deciding factor for selection of appropriate material for preparation of advanced polymer electrolyte.

Table 1. Characteristics of constituents of polymer electrolytes. With permission from Ref. [10] Copyright © 2017 IOP Publishing

| Polymer Host | Plasticizer |
|---|---|
| • Provide fast segmental motion of polymer chain<br>• Low glass transition temperature<br>• High molecular weight<br>• Low Viscosity<br>• High degradation temperature<br>• High Dielectric Constant<br>• Must have electron donor groups | • Low Melting Point<br>• High Boiling Point<br>• High dielectric constant<br>• Low viscosity<br>• Easily Available<br>• Economic<br>• Inert to both electrodes<br>• Good Safety and Nontoxic Nature |
| Solvent | Nanofiller |
| • Abundant in Nature<br>• Non Aqueous in Nature<br>• Low Melting Point<br>• Low Viscosity<br>• Large Flash Point<br>• High Dielectric Constant<br>• Good Solubility for Polymer and Salt | • High Polarity<br>• Low Melting and High Boiling Point<br>• Safe and Nontoxic<br>• Environmental friendly and cost effective<br>• Inert to All Cell Components.<br>• Act as Lewis Acid for Interaction with Polymer<br>• High Dielectric Constant for better dissociation of salt |
| Salt | Nanoclay |
| • Low Lattice Energy for More Availability of Free Ions<br>• High Ionic Conductivity<br>• High Mobility<br>• Broad Voltage Stability Window<br>• Low Ion Pair Formation at High Content<br>• Large Anion Size<br>• Small Cation size for fast migration between the electrodes<br>• High Thermal and Chemical Stability<br>• Large Ion Transference Number<br>• Inert Towards Cell Components | • Layered/unique structure with high aspect ratio (~1000).<br>• Complex rheological behavior<br>• Amorphous behavior with acid base properties<br>• Greater ability for intercalation and swelling<br>• Increase solubility of salts<br>• High swelling index (water and polar solvents)<br>• High cation exchange capacity (CEC) (~80 meq/100 g)<br>• High external/internal surface area (~31.82 $m^2 g^{-1}$)<br>• Appropriate interlayer charge (~0.55)<br>• Adjustable hydrophilic/hydrophobic balance |
| Ionic Liquid | Nanorod/Nanowire/Nanobelt |
| • Good Thermal stability<br>• Wide electrochemical stability<br>• Low Melting Point<br>• Low viscosity for fast transport<br>• Negligible Volatility<br>• Non-flammability<br>• Negligible Vapor Pressure<br>• High Ionic Conductivity<br>• High Polarity<br>• High Dielectric Constant | • High specific surface area<br>• High aspect ratio<br>• Long continuous path<br>• Easily alignment perpendicular to electrodes<br>• Oxygen vacancies on the surface for cation<br>• Less agglomeration at high content<br>• High thermal stability<br>• Better chemical stability<br>• Long cycle stability |

**Polymer nanocomposites**

Polymer nanocomposites (PNCs) are the recently adopted composite polymer electrolyte in the electrolyte community due to various advantages such as high safety, inflammable nature, high reliability and broad thermal/voltage stability. The synthesization of the composite polymer electrolyte include the addition of a salt in the host polymer matrix and a nanoparticle. The most critical requirement with the PNC is the formation of the amorphous content that will improve the electrode/electrolyte interface [12, 18]. The increased amorphous content also improves the use of full electrode material during cell operation. The two parameters ionic conductivity and the cation transference number are generally observed for deciding the electrolyte in energy storage/conversion devices. The conductivity enhancement can be done by various approaches, the addition of nanofiller, nanoclay, nanorod, nanowire. The main characteristics are that the nanoparticle must have high surface area and oxygen vacancies. The former one results in the formation of a long continue conducting path for cation migration while later one provides additional coordinating sites for the cation. In case of nanoclay, the polymer chains get intercalated inside the clay galleries and this increases the interchain separation as well as gallery spacing. This increase in both parameters leads to overall enhancement in the conductivity. Beside the nanofiller, nanoclay now nanorod/nanowire are attractive candidates being used for the preparation of the polymer nanocomposite. The two advantage with them is the formation of a long conductive continuous network and the oxygen vacancies for ion migration. Then one fundamental requirement is that the for ideal electrolyte cation transference number must be unity. So, the man approach is the mobilization of the anion to prevent the concentration polarization at the electrodes. The anion may be covalently bonded to the polymer backbone or some anion acceptors may be used.

## 3. Materials and methodology

### 3.1. Preparation Methods

The preparation of the polymer nanocomposites is done by various methods, such as, Solution Cast Technique, In Situ Polymerization Technique Melt Intercalation Technique, Spin Coating Technique, Hot Press Technique, and Dip Coating Technique.

***Solution Cast Technique***

It is the traditional method due to its ease of fabrication and produces polymer film from various thicknesses (50-300 μm). First of all the specified amount of the polymer host is dissolved in the appropriate solvent by continues stirring. Then the appropriate salt content is added to the homogenous polymer matrix and again stirred till a homogenous and transparent solution is obtained.  Now, to prepare the polymer nanocomposite, firstly the nanoparticle is added in a solvent and ultra-sonication is performed for better dispersion. Then the obtained solution is added to the polymer salt solution and stirred till a homogenous solution is obtained. Finally, the viscous solution is cast in the petri dish and kept at room temperature for evaporation of the solvent. Then kept in a vacuum oven to completely remove the residual solvent and film is peeled off from petri-dish [13]. The obtained film is kept in a desiccator with silica gel for further characterization and prevention from the moisture. For high-quality film, the solvent must be free from water content and should dissolve the polymer, salt, and nanoparticle [25].

***Spin Coating***

This method is almost identical to the solution cast technique. The key advantages of spin coating technique are simplicity and relative ease of preparation and produce uniform films from a few nanometers to a few microns in thickness. In this technique, a small amount of solution is dropped on a substrate and kept on the spin coater which can be rotated at desired sped. The centripetal acceleration spreads the mixture on a substrate and followed by heating to evaporate the solvent. One point that must be taken care of is that the substrate spinning axis must be perpendicular to the substrate which is to be coated. The thickness of the film is influenced by, (i) viscosity of mixture, (ii) concentration of mixture, (iii) speed of rotation, and (iv) spin time. However, this method is advantageous only for low viscosity mixture not for too high viscosity mixture. For a gel mixture, the spin coater rotation is not enough to spread the mixture droplet to form thin film [26-27].

***Hot Press Technique***

Hot press technique is a unique technique with features: low cost, solvent-free, a good film with dense materials and is a fast method. The set-up consists of weighing cylinder, heating chamber, basement, (d) temperature controller. Firstly, the polymer, salt, nanoparticle are grounded in the agate pestle for the required time. Then the obtained mixture is heated (close to melting temperature of polymer) and the obtained slurry is sandwiched between the stainless steel (SS blocks. Then pressure controlled system presses the slurry and obtained film is used for further characterizations [28-29].

***Dip Coating***

The Dip Coating unique feature is that it enables us to obtain high-quality film on both side of the substrate and is low-cost technique. First of all the chosen substrate is dipped in the solvent [30]. The three-step process occurs, immersion, then deposition & drainage, followed by evaporation of the solvent. In immersion, the substrate is immersed in solution at a desired speed that provides sufficient time for coating. In the next step deposition & drainage, the substrate remains dipped in the solvent for sufficient time (dwell time) to enable interaction of the substrate with a solvent. Now, the substrate is pulled out slowly at a constant speed that leads to the formation of a thin film on the substrate. The in the final step evaporation, the solvent is evaporated and the substrate may be heated to completely remove the extra solvent. The thickness and quality of the film can be controlled by the speed of removal and density of the solution.

***Melt Intercalation Technique***

This is an important technique due to advantages such as; environment-friendly, cost-effective and no need of solvent. One point to be noted is that for better dispersion of nanoparticle, the optimization of thermal properties need to be performed as, the high temperature may alter the surface properties of the nanoparticle and may degrade it. First of all, the annealing of the polymer matrix is done at high temperatures, then nanoparticle is added, followed by kneading the composite to achieve uniform distribution. In conclusion, the surface modification of clay/nanofiller, compatibility of filler with host polymer and processing conditions affects the dispersion of nanoparticle [31-32].

### 3.2. Characterization techniques

A battery assembly consists of three components, cathode, anode and electrolyte/separator. The solid polymer electrolyte (polymer nanocomposite) is sandwiched between the two electrodes and plays the dual role of both electrolytes (for ion migration) as well as a separator (for physical separation of electrodes to prevent short-circuit).

So, the electrolyte plays a key role during operation of the battery. So, keeping in mind the desirable device for practical application there are some requirements or performance parameters that need to be measured experimentally in any polymer electrolyte cum separator [33].

*Fourier Transform infra-red spectroscopy (FTIR)*

The salt dissociation capability of the polymer host plays an important role in deciding the suitable polymer electrolyte with balanced properties. The polymer host must have the high dissolution capability to boost the salt dissociation into cation and anion and to fulfill the purpose of the fast ionic conductor. The dissociation capability of the host polymer is indicated by the presence of the polar group of polymer ($-\mathrm{O}-, \mathrm{C} \equiv \mathrm{N}-, \mathrm{C} = \mathrm{O},$) and how strongly it gets coordinated with the cation. Another important point is the less tendency of the chain reorganization so that complete polymer chain can effectively participate in separating the cation and anion. Further the supportive role is played by the salt in deciding the overall solvation ability. The salt must have bulky anion as they have low lattice energy and supports in overcoming the coulombic interaction between ions. The mutual dissolution ability of the polymer and salt lead to the cation migration via the segmental motion of the polymer chains. Beside the polymer chain and the salt solvent also influence the dissolution ability of the all participating species. The solvent must have the high dielectric constant and the high dipole moment so that the ion association can be hindered. The overall effect of the three key players is the effective number of free ion carriers for transport and elimination of the ion pair as they do not participate in the conduction. So, to achieve the balanced properties the suitable combination of the polymer, salt and the solvent is a vital requirement for the fast ionic conductor. The properties of the individual polymer, salt and the solvent are discussed in the forthcoming section.

*Complex Impedance Spectroscopy (CIS)*

As ionic conduction decides the performance of an energy storage system. The ionic conductivity depends on the molecular weight of the polymer, cation/anion size, and viscosity of polymer chains.  The polymer electrolyte, therefore, must possess high ionic conductivity and prevent the flow of the electrons hence the negligible electronic conductivity. As the migration of anions may affect the battery performance so its migration is avoided during the design of any polymer electrolyte system. The main strategy is to immobilize the anion and the only cation dominates in the conduction. The ionic conductivity of any polymer electrolyte system (σ) is linked with two parameters; a number of free charge carriers and ion mobility (equation 1).

$$\sigma = \sum n_i q_i \mu_i \qquad (1)$$

Here, Here, $n_i$ represents the effective number of charge carriers of type i; $q_i$ is the charge of the charge carriers, and $\mu_i$ is the mobility.

An important factor that influences the ionic conductivity is the activation energy. This is the minimum energy that is required for ion migration and is obtained from the conductivity vs. temperature plot [34]. Activation energy decreases with the increase of the temperature. Generally, it is supposed that the crystalline phase dominates below the melting temperature while the amorphous phase is effective above the melting temperature. So, both the regions are examined by two conduction mechanisms dependent on temperature, (i) Arrhenius behavior and (ii) Vogel-Tammann-Fulcher (VTF) behavior. The Arrhenius equation describes the relation between $\log \sigma$ and $T^{-1}$, as shown in Eq. (2)

$$\sigma = \sigma_o\ exp\ \left(-\frac{E_a}{kT}\right) \qquad (2)$$

Here, $E_a$ is the activation energy, which can be calculated from the nonlinear least-squares fitting of the data from plots of log $\sigma$ vs. $T^{-1}$. For polymer electrolytes, plots of $\sigma$ vs. $T^{-1}$ are typically nonlinear, indicating that the conductivity mechanism involves an ionic hopping motion coupled with the relaxation and/or segmental motion of the polymeric chains.

The VTF equation can be derived from the quasi-thermodynamic models with the free volume and configurational entropy, and its behavior can be related coupled motion with the segmental motion. This can be expressed by equation 3.

$$\sigma = \sigma_o T^{-1/2} \exp\left(-\frac{B}{T - T_o}\right) \qquad (3)$$

Here, $\sigma_o$ is the pre-exponential factor, which is related to the number of charge carriers $n_i$, B is the pseudo activation energy of the conductivity, and $T_o$ is the reference temperature associated with the ideal glass transition temperature (Zero mobility temperature). For practical aspects, the high ionic conductivity of the order of $10^{-2}$-$10^{-3}$ S $cm^{-1}$ is required and electronic conductivity of the order of $10^{-6}$-$10^{-10}$ S $cm^{-1}$. This enhances the charging/discharging rate and cyclic stability.

Table 2. Definition and comparison of the laws governing the transport phenomena in polymer electrolytes. With permission from Ref. [35] Copyright ©1997 Elsevier.

| | Arrhenius | VOGEL-TAMMAN-FULCHER (VTF) | WLF (WILLIAMS, LANDEL & FERRY) |
|---|---|---|---|
| Expression | $\sigma = \sigma_o\ exp\ \left(-\frac{E_a}{kT}\right)$ | $\sigma = \sigma_o T^{-1/2} \exp\left(-\frac{B}{T - T_o}\right)$ | $\sigma = \sigma_o \exp\left(-\frac{(C_1(T - T_{ref})}{C_2 + \mathrm{T} - T_{ref}}\right)$ |
| Typical Values | – | $\sigma_o = 0.4$ S cm-1, B=2.210$^{-3}$ K-1, To=210, K≈Tg-50 | $C_1 = 17.4$, $C_2 = 52\ K$, $T_{ref} = T_g = 240\ K$ |
| Derived From | $k = \mathrm{a}\ exp\ \left(-\frac{E_a}{kT}\right)$ | Variation of viscosity with T $\eta \propto \mathrm{T}^{1/2} \exp\left(-\frac{\mathrm{B}}{\mathrm{T} - \mathrm{T_o}}\right)$ | Scaling factor $a_T$ for relaxation: $log(a_T) = -\frac{(C_1(T - T_{ref})}{C_2 + \mathrm{T} - T_{ref}}$ |
| Valid Relations Needed | y=mx+c | $D = \frac{kT}{6\pi\eta r}$ Stokes-Einstein | $\sigma = \frac{Ze^2}{kT} D$ Nernst-Einstein |
| Correspondance VTF↔WLF | $T_o \to 0$ to in VTF | Neglect $T^{1/2}$ prefactor in $\sigma_o$ from VTF $C_1C_2$=B | $C_2 = T_{ref} - T_o$ |

*i-t characteristics (Transference Number)*

It is one of essential parameters that must be close to the unity (~1) for a good ionic conductor and function of the ionic conductivity. One main drawback associated with the polymer electrolytes is concentration polarization (CP) on the electrode interface which hinders the ion migration during cell operation. This affects the overall performance of the device and observed in terms of the low energy density and power density. So, the main strategy to reduce the CP is to immobilize the anion and keep only the active species as the main hero of the charge transport scenario. This fulfills the idea of the single state ionic conductor. So, the large value of the cation transference number also plays an effective role in the enhancement of the fundamental property that is ionic conductivity. Two approaches are adopted for immobilizing the anion. One approach suggests the anion interaction with the hydrogen of the polymer backbone (hydrogen bond formation) another is the introduction of the anon receptors. Generally, there are two terms here, cation transference number and ion transference number.

Table 3. Definition and comparison of the transference number in polymer electrolytes

| | Cation transference number | Ion transference number |
|---|---|---|
| Expression | $t_{cation} = \left(\frac{I_s(V - I_oR_o)}{I_o(V - I_sR_s)}\right)$ | $t_{ion} = \left(1 - \frac{I_e}{I_t}\right)$ |
| Symbol Depiction | V is the applied dc voltage for sample polarization, $I_o$ and $I_s$ are the currents before and after polarization, $R_o$ and $R_s$ are the initial and steady-state resistance of the passivation layers | $I_t$, is total initial current due to ions and electrons contribution and $I_e$ is residual current due to electrons contribution only |
| Value | ~1 | ~100 % |

*Differential Scanning Calorimetry (DSC)*

Glass transition temperature ($T_g$) is one of the most fundamental properties of the polymer electrolytes that is directly linked to the electrochemical properties. It is influenced by the polymer chain arrangement, crystallinity, viscosity, polymer interactions or polarity and molecular weight of the polymers. As most of the polymers are crystalline in nature. The ionic conduction in case of the polymers is supposed to occur via the amorphous phase and segmental motion of the polymer chains which is further linked with the flexibility of chains. This flexibility criterion is specified by the glass transition temperature ($T_g$) value. It is defined as a transition temperature at which any system goes from rigid to rubbery/viscous phase. Below $T_{g,}$ there is no ion migration or chain are not moving. While above the $T_g$ there is a drastic change in the intrinsic properties (density, specific heat, mechanical modulus, mechanical energy absorption, their dielectric and acoustical equivalents) of the polymer that supports the fast segmental motion or less viscosity [36]. The differential scanning calorimetry (DSC) measurement is performed to measure the $T_g$ and it also provides the melting and crystallinity of the used material. The lower value of the $T_g$ indicates enhanced flexibility for the polymer matrix. This increase in flexibility improved the ion swimming rate between electrodes with the coordinating sites of the polymer chains. So, the main strategy for lowering the value of the $T_g$ is to alter the polymer chain arrangement and disruption of the covalent bonding between the polymer chains. This can be done by the addition of nanofiller, plasticizer, nanoclay etc. Generally, the addition of the above said particles increase the free volume available for the ion migration and this makes faster ion migration.

*Crystallinity*

The ordering of the polymer chain (long range and short range) affects the ion migration in a polymer electrolyte. As, long range order is associated with the crystallinity which in case of the polymer electrolyte's must be lower for faster ion migration. So, the crystallinity of any system provides sufficient information regarding the material and how it will play its role during the ion transport. The crystallinity of any polymer material is obtained by the X-ray diffraction (XRD) and the Differentials scanning calorimetry (DSC). It needs to be mentioned here that the amorphous content supports the fast ion migration in case of the polymer electrolytes. The amorphous content is achieved by disrupting the crystalline arrangement of the polymer chains. The best approach is the addition of the nanofiller and alters the arrangement by the Lewis acid-base interaction with polymer chains. The intercalation of the polymer chains in the clay gallery also lowers the crystallinity. This increases the free volume available for the ion migration as now segmental motion of the polymer chains becomes faster and ion jump faster from one coordinating site to another.

The amorphous content can be visualized by the Field emission scanning electron microscope (FESEM), Transmissions electron microscope (TEM) and atomic force microscope (AFM).

*Linear sweep voltammetry (LSV)*

For fulfilling the requirement of polymer electrolyte for the application aspects it must have a broad electrochemical stability window (ESW) and depends on the stability of individual cation/anion. It is defined as the difference between the potentials of the oxidation reaction and reduction reaction. As one characteristic a polymer electrolyte must possess is that it must be inert toward both electrodes. For this, the oxidation potential must be higher than the embedding potential of cation ($Li^+$) in the cathode and the reduction potential must be lower than that of lithium metal in the anode. The desirable range of the ESW is ~4-5 V and is enough for the commercialized systems.

*Thermogravimetric analysis (TGA)*

Thermal stability of a battery device is important to avoid the material decomposition and explosion during cell operation. So, the thermal stability is investigated by TGA to check the safety window of a polymer electrolyte. As, different material have different decomposition temperature range and for long-term cycle stability all material must lie beyond the decomposition range. So, for the safe working of the battery system, the thermal stability must be large. The approaches for the enhancement of the thermals stability is the addition of nanoparticles in the polymer matrix.

*Stress-Strain curve*

During the commercialization of the battery or any energy storage device, one important property is the mechanical stability. So, the along with high cation transference number and the ionic conductivity it must have mechanical property sufficient enough for long term cycle stability. The mechanical stability is measured by the stress-strain curve in terms of the modulus, stress, and strain. As during the cell packaging the polymer electrolyte is sandwiched between the two electrodes so, it must have the capability to sustain the pressure applied. The polymer electrolyte having poor mechanical strength may short-circuit the electrodes during cell operation. So, the material must be flexible by nature and brittleness is avoided. It must have inherent characteristics that may absorb the small internal shocks and does not affect the electrodes. There are various strategies that are adopted for the enhancement of the modulus, stress, and strain. The most suitable and the attractive approaches are the dispersion of the nanofiller, nanoclay and nanoparticle it high aspect ratio such as nanorod, nanowire, nanotube. Here these nanoparticles may play the role of the crosslinking center and overall enhancement in the mechanical property of the complete polymer matrix is achieved.

## 4. Recent Updates

### 4.1. Nanofiller dispersed polymer nanocomposites

There are various reports toward the addition of the different nanofiller in the polymer matrix for improving the electrical, thermal and mechanical properties. Generally, all nanofiller addition suppresses the crystallinity and crosslinking alters the polymer chain arrangement. Another effective point is that nanofiller supports salt dissociation via Lewis acid-base interactions and provides additional conducting sites for ion transport. But still, there is lack of availability of a polymer matrix with enhanced amorphous content and weak polymer-nanofiller interaction results in nanofiller agglomeration. So, Lin et al., [37] reported the preparation of (PEO)−monodispersed ultrafine $SiO_2$ ($MUSiO_2$) composite polymer electrolyte (PEO-$MUSiO_2$ CPE) via in situ hydrolysis of tetraethyl orthosilicate(TEOS) in PEO solution. Basically, advantage with the in situ hydrolysis is that the polymer-nanofiller interaction is more effective in suppressing the crystallinity by linking the polymer chains with the nanofiller surface. Two possible dominant interactions exist here, (i) between the hydroxyl groups at the ends of PEO chains with the surface of $SiO_2$, (ii) wrapping and embedding of PEO chains inside $SiO_2$ spheres.

TEM image evidence that the higher polymer density was for the in situ prepared composites as compared to ex-situ and indicates the presence of strong interaction between the polymers a nanofiller (Figure 6 d-f). Further, XRD also evidences the superior evidence of reduction of crystallinity as compared to the ex situ synthesis (Figure 6 j). Further FTIR and DSC also evidence the reduction of crystallinity by in situ synthesis. One reason is disruption of the polymer chain reorganization tendency due to the interaction between MUSiO2 spheres and polymer chains. Another reason is improved surface area due to uniformity in size and distribution.

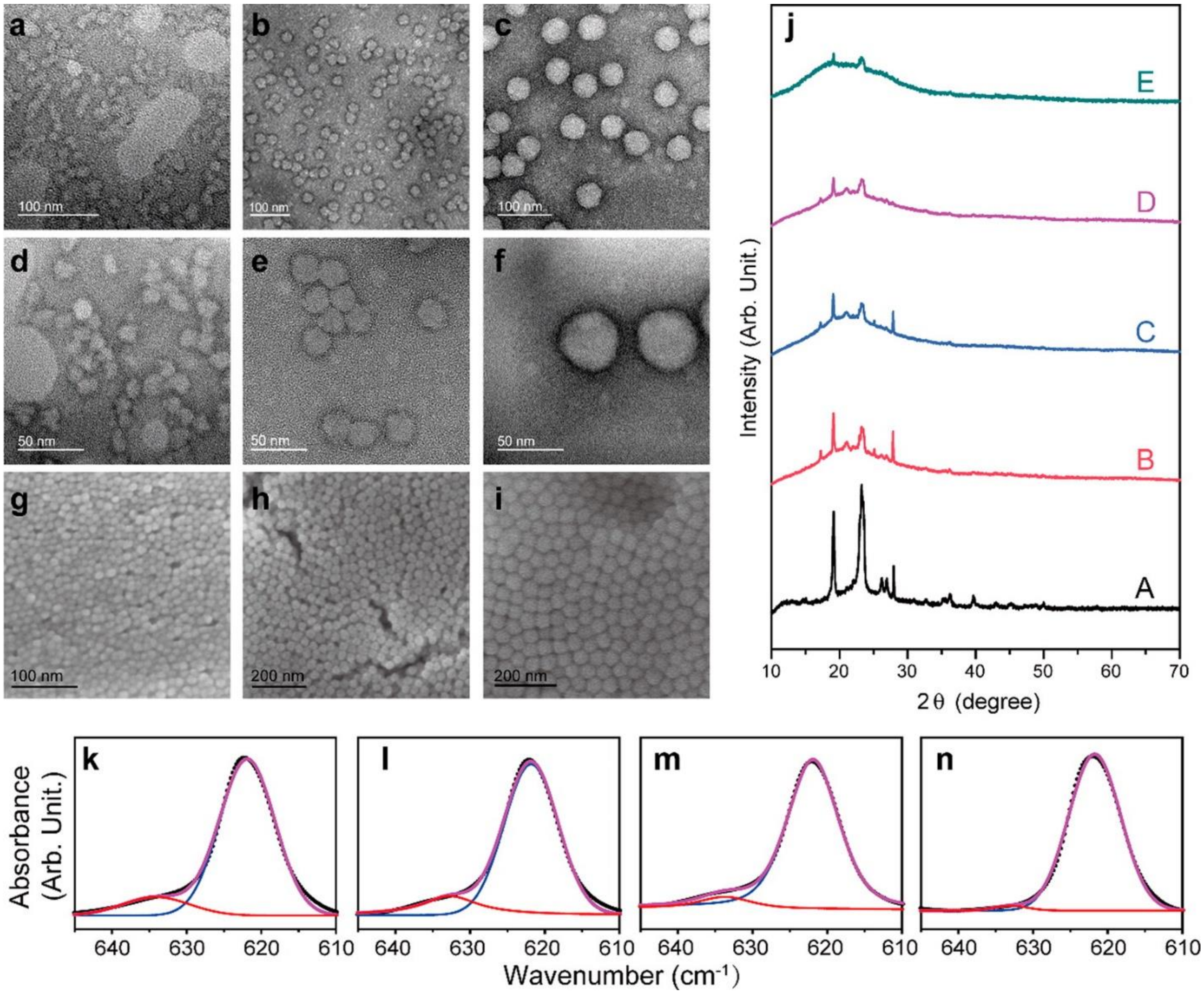

Figure 6. Characterizations of in situ CPE. (a−f) TEM images of in situ PEO-MU$SiO_2$ composite with different sizes of ~12 nm (a, d), ~ 30 nm (b,e), and ~45 nm (c, f). PEO was stained with 0.1% phosphotungstic acid to show better contrast. (g–i) SEM images of as-synthesized correspondingMU$SiO_2$ spheres (without PEO) with various sizes of ~12 nm (g), ~ 30 nm (h), and ~45 nm (i). (j) Comparison on XRD spectra of pure PEO(A), ceramic-free SPE (B), PEO-fumed SiO2 CPE (C), ex situ CPE (D), and in situ CPE (E). (k−n) FT-IR spectra at 610−645 $cm^{-1}$ and corresponding Gaussian−Lorentzian fitting of the $ClO_4^-$ absorbance for ceramic-free PEO SPE (k), PEO-fumed $SiO_2$ CPE (l), ex situ CPE (m), and in situ CPE (n). With permission from Ref. [37] Copyright © 2016 American Chemical Society.

Another remarkable evidence is increased number of free ion charge carriers (left to right) in figure 6 k-n. The degree of dissociation of salt was higher for the sample synthesized by in situ synthesis following the relation in situ (98.1 %) > ex situ (92.8 %) > PEO-fumed $SiO_2$ CPE (87.4%) > ceramic-free SPE (85.0%). This was attributed to the simultaneous achievement of two parameters, uniform distribution of $SiO_2$ and increased the segmental motion of polymer chains. The ionic conductivity also shows enhancement for polymer electrolyte synthesized by in situ and is in the range $10^{-4}$-$10^{-5}$ S $cm^{-1}$ (at ambient temperature) and $1.2\times10^{-3}$ S $cm^{-1}$ (at 60 $^oC$). Further, the electrochemical stability window was improved for in situ synthesis (> 5.5 V) as compared to ex-situ synthesis (~4.7 V) and may be due to the strong adsorption effect on anion for in situ synthesis [38]. Further the rate capability test of solid-state battery (LFP/CPE/Li) displays double capacity retention for in situ synthesis (120 mAh/g at 90 $^oC$ and 100 mAh/g at 60 $^oC$) as compared to ex situ (65 mAh/g), while for ceramic free it was 50 mAh/g. For in situ cycle stability was achieved and after 80 cycle capacity delivered was 105 mAh/g. After 80 cycles decrease was observed due to the poor interface stability and dendrite growth may occur.

Pal et al., [39] reported the PMMA-$LiClO_4$ based polymer nanocomposite electrolyte with $TiO_2$ as nanofiller using solution cast technique. XRD diffractograms depict complete dissociation of the salt and PMMA peak broadens with the addition of nanofiller. The addition of 1 wt. % $TiO_2$ evidences the optimum enhancement of the amorphous content. Further TEM analysis evidences the non-uniform distribution and clustering, with no effect on particle size with nanofiller loading.

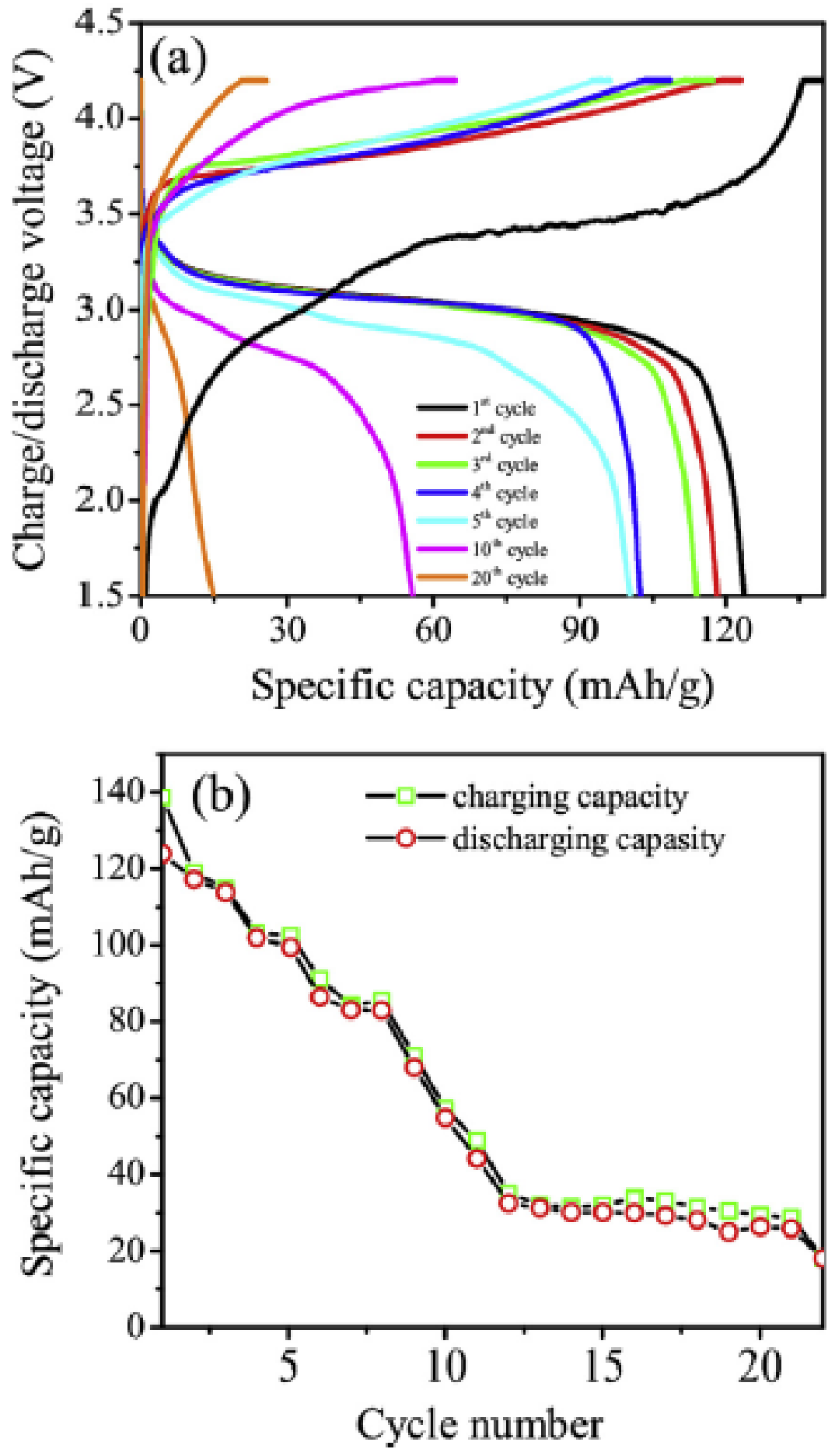

Figure 7. (a) Charge-discharge profile at C/16 of graphite/plasticized PMMA-$LiClO_4$-1 wt% $TiO_2$/$LiCoO_2$ lithium ion polymer coin cell at 25 $^oC$. (b) Cycling performance of plasticized PMMA-$LiClO_4$-1 wt% $TiO_2$ electrolyte at C/16 at 25 $^oC$. With permission from Ref. [39] Copyright © 2018 Elsevier.

The changes in the peak of pure PMMA (1732, 1492, 1444, 1385, 1192, 1150, 988, 966, 844 and 756 $cm^{-1}$, C=O asymmetric stretching of the carbonyl group, –$CH_2$ scissoring, O-$CH_3$ bending, -$CH_2$ twisting, C-O-C bending, carboxylic acid ester group, C-C symmetric stretching, -$CH_2$ wagging, -$CH_2$ asymmetric rocking and –$CH_2$ rocking) with addition of the salt and nanofiller in FTIR spectra confirms the presence of polymer-ion, ion-ion interactions. Also, the number of free charge carriers were more for low clay content and attributed to the cation interaction with electron rich group of PMMA. DSC analysis shows a decrease of the $T_g$ with the addition of nanofiller while at high nanofiller content increase is due to nanofiller cluster formation. TGA analysis shows thermal stability up to 250 $^oC$ and is in the desirable range for application purpose. The solid-state battery configuration with PMMA-$LiClO_4$- 1 wt% $TiO_2$ as electrolyte shows the discharge capacity of 123 mAh/g for the 1st cycle and 136 mAh/g for 1st charging cycle with coulombic efficiency 100 % (Figure 7).

As a lot of reports are available by addition of nonporous inorganic nanofiller but mesoporous nanoparticle may be vital alternative and may fulfill the dream of single ion conductor. So, the suppression of crystallinity along with immobilization of anion are simultaneously resolved by introducing the bulky imide group. Liang et al., [40] prepared the PEO-PMMA based polymer nanocomposites with two salts ($LiClO_4$ or LiTFSI) and nano-Al2O3 was used as nanofiller by solution cast technique. FESEM analysis showed the more uniform dispersion for PEO-PMMA-LiTFSI$Al_2O_3$ and a more even surface morphology which evidences the lowering of the interfacial resistance. Fig. 8 a shows the impedance spectrum for the polymer-based electrolyte prepared at room temperature. The conductivity was increased with the nanofiller addition and is about $9.39\times10^{-7}$ S $cm^{-1}$. This increase in the conductivity was due to the increased salt dissociation and increased flexibility. The thermal stability displayed by the TGA graph was above 300 $^oC$ and was improved after addition of the nanofiller. Also, the higher stability was shown with LiTFSI salt as compared to the $LiClO_4$. The temperature dependence of the ionic conductivity follows Arrhenius behavior and activation energy decreases from the 20.10 kJ/mol to 10.96 kJ/mol. The voltage stability window of the prepared system was up to 4.9 V. Further mechanical analysis was analyzed from the stress-strain curves. The tensile strength was 2.84 MPa (PEO-PMMA-$LiClO_4$-$Al_2O_3$) with an elongation-at-break value at 31.7 %. While for PEOPMMA-LiTFSI-$Al_2O_3$ electrolyte the tensile strength increased to 3.26 MPa (elongation-at-break value at 11.7%.) and is higher than the PEO-PMMA-LiTFSI (tensile strength= 2.78 MPa, elongation-at-break value at 9.5%).

Kim et al., [41] reported the preparation of polymer nanocomposite using the functionalized mesoporous silica (FMSTFSISPE) nanoparticles (av. size 50 nm) with PEO as host matrix. The prepared films were transparent and display smooth surface morphology with optimum content 30 wt. %. DSC analysis evidences the single glass transition temperature and absence of melting peak suggest its use in a wide range. Also, for 1 wt. % loading it shows a comparatively better effect in suppressing the crystallinity as compared to nanofiller and may be due to the uniform distribution which lowers the reorganization of polymer chains. AT high content increase in the $T_g$ was observed and was attributed to the formation of temporary crosslinking due to the high surface area and mobility reduces. The highest ionic conductivity was ~$10^{-3}$ S $cm^{-1}$ (At 25 $^oC$) for 30 wt % FMS-TFSISPE ($E_a$=26 kJ/mol) and was 10 times higher as compared to desirable limit (Fig 8 b). This was improved than the nonporous silica which displays conductivity ~ $2\times10^{-5}$ S $cm^{-1}$ (At 25 ºC) ($E_a$=34 kJ/mol). Also, the $Li^+$ transference number ($t_{Li}^+$) was ~0.9 and evidences the single-ion conductive matrix. This increase in conductivity and $t_{Li}^+$ with the mesoporous silica and porous nature along with large charge carriers overall supports the fast ion conduction. Further mechanical properties were improved and modulus was $3\times10^4$ Pa for 30 wt % FMS-TFSISPE and follows a relative trend with $T_g$. One attention-grabbing point was that both modulus and conductivity were improved simultaneously. It was concluded that the well-ordered mesoporous channel (high surface area and high pore volume) in the FMS-TFSISPE nanoparticles are superior's candidate for fulfilling the criteria of single ion conductor where anion is in an immobilized state in the pore wall.

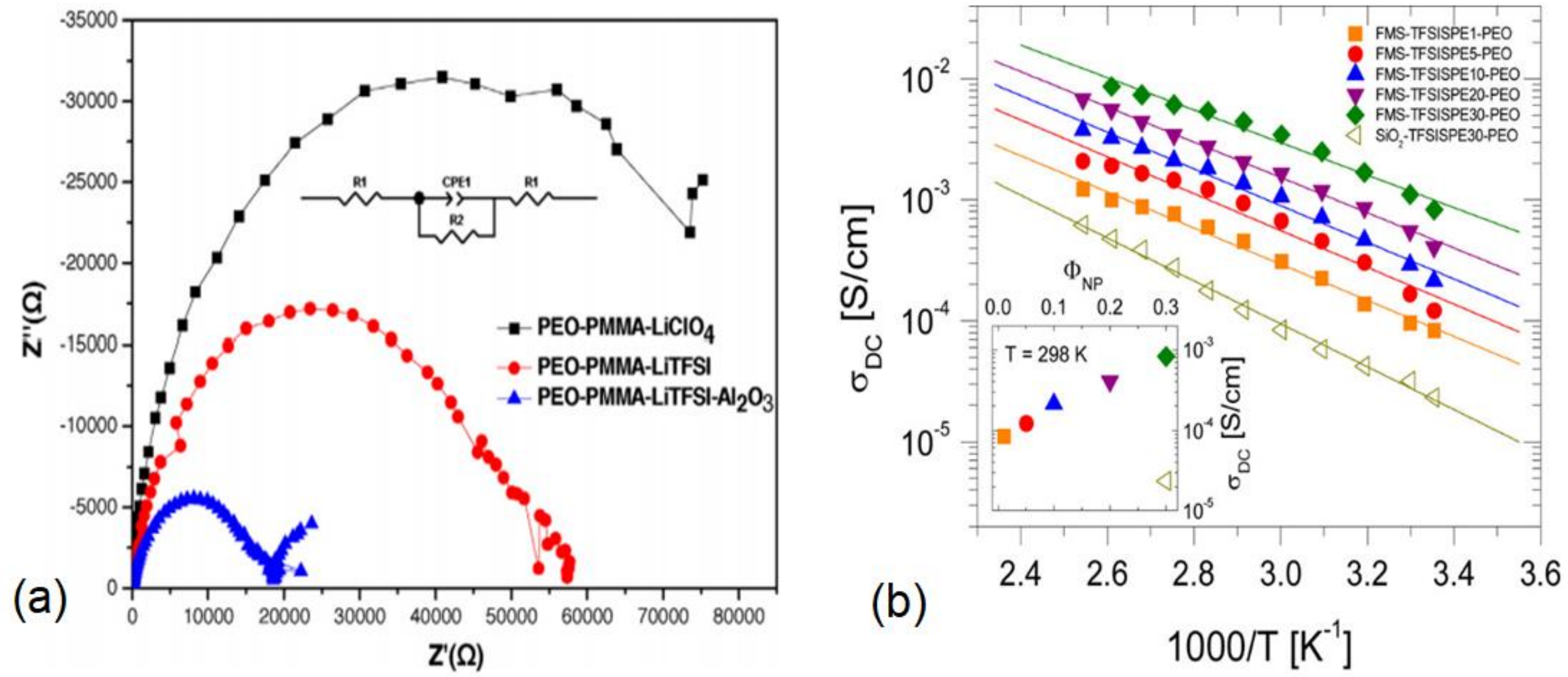

Figure 8. (a) Impedance spectrum for the SS/PEO-PMMA-lithium salt (EO/$Li^+$ = 20)/SS cell at room temperature. With permission from Ref. [40] Copyright © 2015 Elsevier. (b) Temperature dependence of the ionic conductivities $\sigma_{DC}$ of the nanohybrid electrolyte series (FMS-TFSISPE-PEO, filled symbols) with increasing amount of FMS-TFSISPE nanoparticles, $\Phi_{NP}$, compared with $\sigma_{DC}$ of the nonporous silica nanoparticles-based electrolytes ($SiO_2$-TFSISPE30-PEO, open symbols) ($\sigma_{DC}$ at 298 K vs $\Phi_{NP}$ in the inset). With permission from Ref. [41] Copyright © 2017 American Chemical Society.

Tang et al., [42] reported the preparation of polymer nanocomposite by dispersing the hybrid nanofiller (montmorillonite clay−CNT hybrid fillers) into PEO-$LIClO_4$ matrix. XRD analysis demonstrates the disruption of the crystallinity with the addition of nanofiller and may be due to the alteration in polymer chain arrangement. The FTIR evidenced the presence of strong interaction between the polymer matrix and cation and addition of hybrid nanofiller support in the smooth migration of the cation. The FTIR deconvolution evidence that for 10 wt. Clay-CNT highest number of free charged were available for condition (Figure 9 a). This increase may be due to the negative surface charge layer on CNT and ether group which increases the salt dissociation rate (Figure 9 b). Another aspect is that there may be an increase of free volume due to less possibility of chain reorganization. This increase in free volume is linked with faster ion mobility as evidenced by the impedance analysis in terms of conductivity. The highest ionic conductivity was $2.07\times10^{-5}$ S $cm^{-1}$ for the optimum 10 wt. % Clay-CNT and with further increase hybrid nanofiller aggregation occurs which lowers the conductivity value. The mechanical strength and elongation were also much higher as compared to pure PEO. This may be attributed to the principal characteristics of the hybrid nanofiller, (i) high aspect ratio and (ii) rough surface. Both parameters lead to an improved interface between the polymer and nanofiller and hence the improved mechanical property.

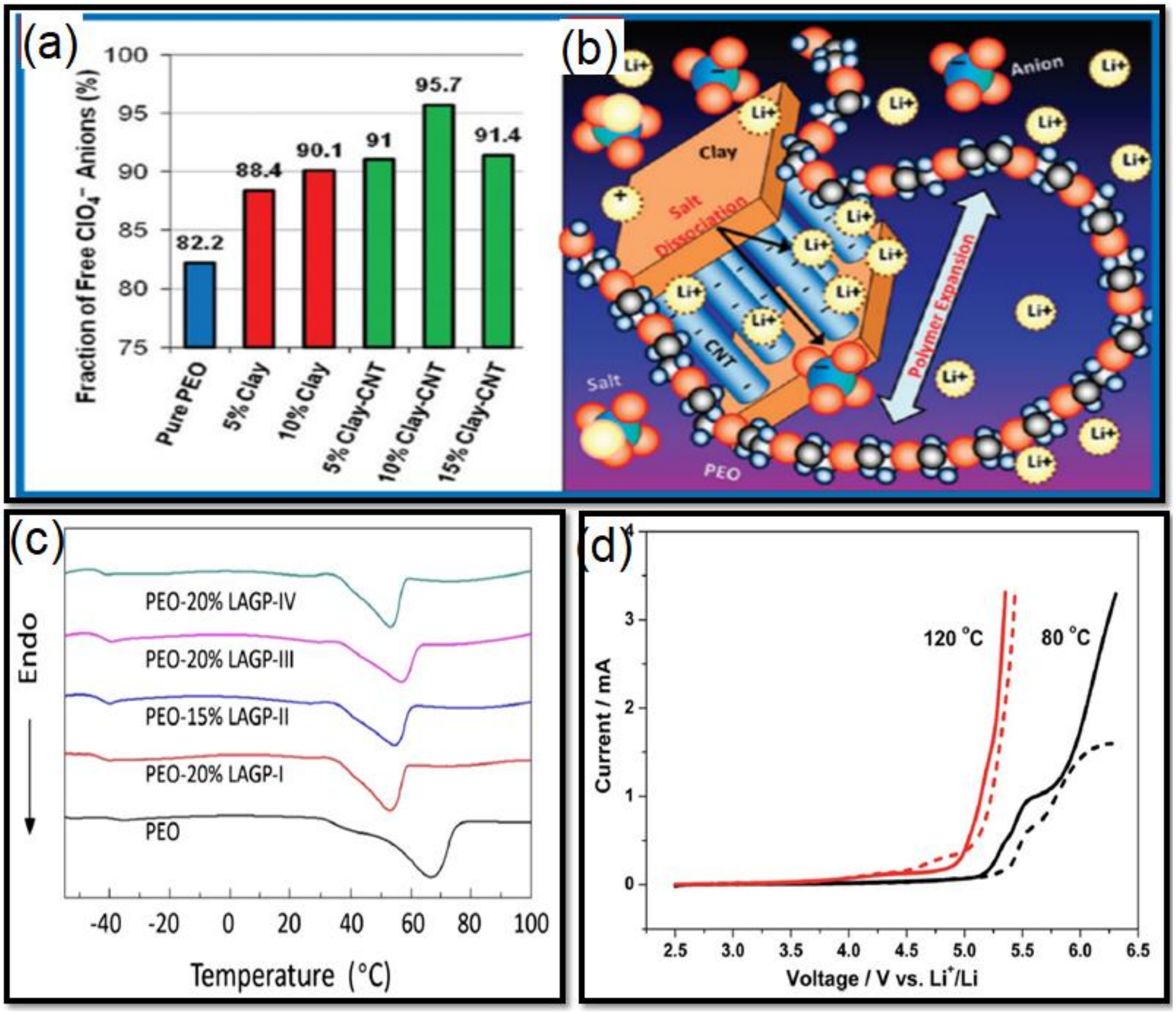

Figure 9. (a) Fraction of dissociated salt ions ($ClO_4^-$ anions) based on FTIR analysis of pure and filled PEO electrolyte. (b) Schematics of the interactions between clay, carbon nanotubes, polymer chains, and lithium salt ions. With permission from Ref. [42] Copyright Copyright © 2012 American Chemical Society. (c) DSC results for the electrolyte membranes. With permission from Ref. [43] Copyright Copyright © 2016 Elsevier., (d) Linear sweep voltammograms of SS/PEO-LiTFSI/Li (solid line) and SS/PEO-MIL-53(Al)-LiTFSI/Li (dotted line) batteries at 80 ºC (black) and 120 ºC (red). The electrolytes were swept in the potential range from 2.5 V to 6.5 V (vs. Li/Li+) at a rate of 10 mV s-1. With permission from Ref. [44] Copyright © 2014 Royal Society of Chemistry.

Zhao et al., [43] reported the preparation of the composite polymer electrolyte based on PEO- LiTFSI, and $Li_{1.5}Al_{0.5}Ge_{1.5}(PO_4)_3$ (LAGP) as nanofiller. The highest ionic conductivity was $6.76 \times 10^{-4}$ S cm$^{-1}$ (at 60 °C) for LAGP-I and it also has smaller particle size. Further DSC analysis shows the lowering of the glass transition temperature and melting temperature (-42 ºC and 52.8 ºC) this indicates that the increased amorphous content favors faster ion transport (Figure 9 c). The cation transference number was > 0.36 and is attributed to the anion blockage.

Fillers also plays the role of the crosslinking network and increase the cation transference number by providing additional conducting pathways. The voltage stability window was 5.3 V for 20 wt. % LAGP-I. The electrochemical analysis of the LiFePO4/PEO-20% LAGP-I/Li cell demonstrates capacities of 166, 155, 143 and 108 mAh $g^{-1}$ at current rates of 0.1, 0.2, 0.5 and 1C, respectively with capacity retention of cell 44% (after 50 cycles).

Zhu et al., [44] reported the preparation of solid polymer electrolyte using the Metal-organic framework aluminum 1,4-benzenedicarboxylate (MIL-53(Al)) is used as a filler with PEO as host polymer and LiTFSI (EO : Li ratio=10, 15, 20, 25) as salt. SEM analysis suggests the uniform surface and morphology was unaltered with cylindrical particles. The ionic conductivity was 10 wt. % MIL-53 (Al) content for EO:Li ratio of 15 : 1 with value $1.62\times10^{-5}$ S $cm^{-1}$ (at 30 $^{o}$C) and $9.71\times10^{-4}$ S $cm^{-1}$ (at 80 $^{o}$C). This was also further supported by the lowering of phase transition temperature from 56.9 $^{o}$C to 50.3 $^{o}$C with the addition of MIL-53(Al) and this enhances the amorphous content and hence the improved conductivity. The $Li^{+}$ transference number was increased from 0.252 to 0.343 with MIL-53(Al) addition and may be due to the formation of the metal-organic framework which enhances the ion mobility. Then the zeta electric potential measurement shows the pH about 7 and it indicates that the MIL-53(Al) particles have strong Lewis acidic properties. This helps in the salt dissociation while anion is coordinated with the Lewis acidic surface of the nanoparticle. The overall effect is the disruption of the crystalline nature and improved ionic conductivity. The electrochemical stability window of the polymer electrolyte was 5.31 V at 80 $^{o}$C and 5.10 V at 120 $^{o}$C with nanoparticle and is larger than the nanoparticle free system which shows 5.15 V at 80 $^{o}$C and 4.99 V at 120 $^{o}$C (Figure 9 d). The thermal stability analysis evidences the thermal stability of about 200 $^{o}$C with the first degradation beginning from 195 $^{o}$C (decomposition of PEO) followed by degradation at 375 oC (decomposition of LiTFSI) [45]. Also, the mechanical properties were enhanced with nanoparticle addition and may be due to the formation of crossing-linking centers for PEO. Then the cyclic performance of the solid-state battery (LiFePO4/PEO-MIL-53(Al)-LiTFSI/Li) was tested and the initial discharge capacity was 127.1 $mAhg^{-1}$ (at 5 C and 80 $^{o}$C) and 136.4 mA h $g^{-1}$ at 120 $^{o}$C. After the 300 cycles, the discharge capacity was 116.0 mAh $g^{-1}$ at 80 $^{o}$C and 129.2 mA h $g^{-1}$ at 120 $^{o}$C. Even after 14 cycles the retention ratios of 52.4 % and 61.3 %, respectively were achieved at 80 $^{o}$C and 120 $^{o}$C.

Vignabooran et al., [46] reported the preparation of the composite polymer electrolyte based on PEO-LiTf-EC-TiO2. The purpose of this research was to study the combined effect of the plasticizer and the nanofiller, as both plays influence the polymer matrix in a different manner. The highest ionic conductivity was $4.9\times10^{-5}$ S $cm^{-1}$ for the 10 wt. % $TiO_2$ (at 30 $^{o}$C, $E_a$=78.8 kJ/mol) and was attributed to the Lewis acid base character of the nanofiller surface. The nano filler surface affects in two ways, one is a reduction of the polymer recrystallization tendency and another is an increase of salt dissociation or lowering in ion pairing. Both simultaneously supports the fast ion migration. Further addition of the EC increases the conductivity upto $1.6\times10^{-4}$ S $cm^{-1}$ for the 50 wt. % EC (at 30 $^{o}$C, $E_a$=57.5 kJ/mol). Further DSC analysis supports the enhancement in the conductivity as both glass transition temperature (-46 $^{o}$C to -50 $^{o}$C) and the melting temperature (60 $^{o}$C to 50 $^{o}$C) show decrease in the signal temperature with addition of the EC in $TiO_2$ based polymer matrix.

Another report by Klongkan et al., [47] investigated the effect of PEG-DOP plasticizer and $Al_2O_3$ nanofiller on the PEO-$LiCF_3SO_3$ based polymer matrix. DSC spectra's shows the decrease of the crystallinity from 37.31 % (PEO-15 wt.%$LiCF_3SO_3$) to 23.11 % (PEO-15 wt.% $LiCF_3SO_3$-20 wt.%DOP) and 18.61 % (PEO-15 wt.%$LiCF_3SO_3$-20

wt.%Al2O3). The decrease in crystallinity on the addition of salt was due to the cation coordination with the PEO and it alters the polymer chain arrangement, hence the increased segmental mobility. Addition of nanofiller increases slightly the $T_g$ and $T_m$ and that may be due to the crosslinking of the polymer chain while crystallinity reduction is observed [48]. The ionic conductivity increases from the 1.00×10$^{-6}$ S cm$^{-1}$ to 7.60×10$^{-4}$ S cm$^{-1}$ (PEO-15 wt. % $LiCF_3SO_3$-20 wt.% DOP) and 8.64×10$^{-5}$ S cm$^{-1}$ (PEO-15 wt.% $LiCF_3SO_3$-20 wt.%$Al_2O_3$). This increase in the conductivity was attributed to increased amorphous phase with the addition of the nanofiller and the plasticizer. The further mechanical analysis supports the both DSC and the conductivity data. The stress at maximum load, percentage strain at maximum load and Young's modulus are 4.4 MPa, 3351% and 15MPa, respectively for the polymer matrix (PEO-15 wt% $LiCF_3SO_3$). This decrease was attributed to the disruption of the polymer chains sliding. Further addition of the DOP also lowers the mechanical properties and that may be attributed to the effective role of the plasticizer which lowers the friction (increase the mobility) and it lowers the mechanical properties [49-50].

Another report using the hybrid nanofiller was by Polu et al., [51] and it was based on the effect of polyhedral oligomeric silsesquioxane-polyethylene glycol (POSS-PEG(n = 4)) nanofiller on the physicochemical and electrochemical properties of PEO-LiDFOB based nanocomposite solid polymer electrolyte. FESEM micrographs depict the change of morphology from rough (of PEO) to smoother on the addition of nanoparticle and may be associated with the complete dissociation of both salt and nanofiller. XRD analysis confirms the insertion of nanofiller in the polymer salt matrix and enhancement of amorphous content is achieved. Further DSC analysis confirmed the increased amorphous content and fast segmental motion. The highest ionic conductivity was 7.28×10$^{-5}$ S cm$^{-1}$ for 40 wt. % of the nanoparticle.

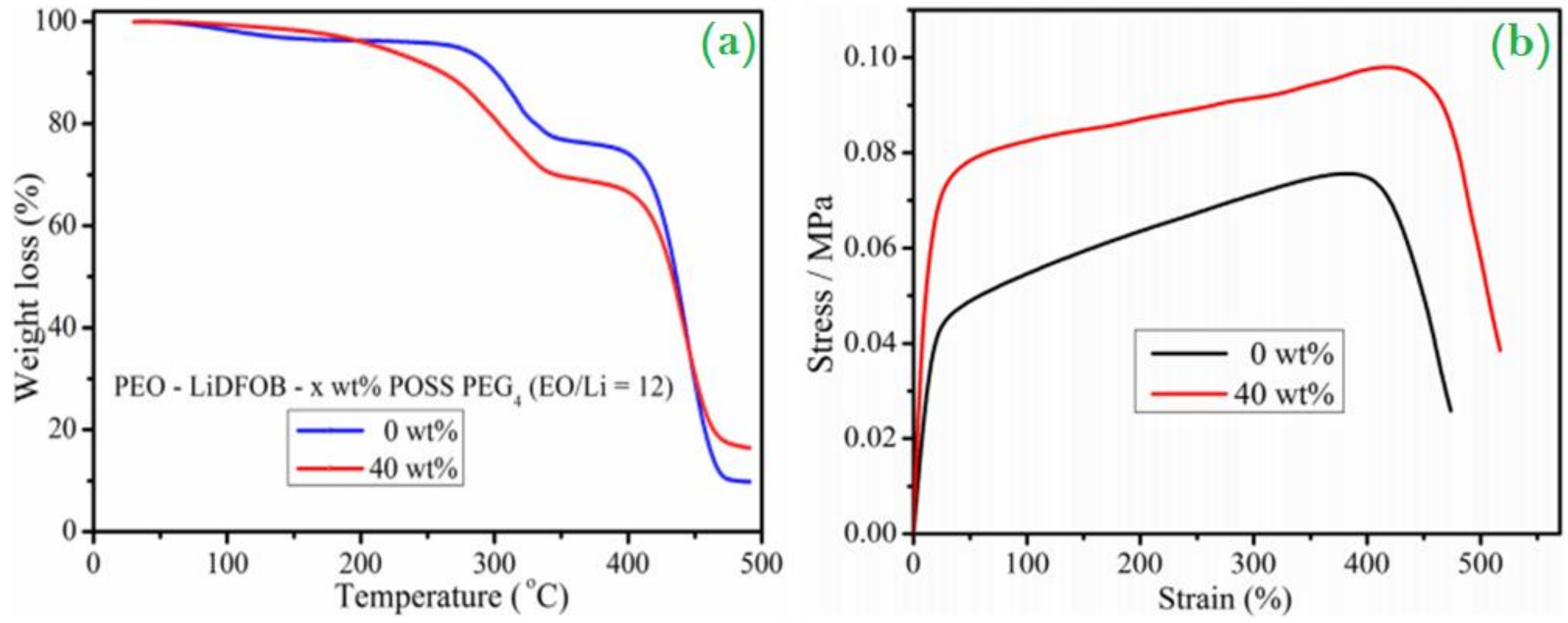

Figure 10. (a) TGA heating traces 0 wt. % and 40 wt. % POSS-PEG doped $PEO_{12}$:LiDFOB electrolyte membranes and (b) Stress–strain curves of $PEO_{12}$:LiDFOB and $PEO_{12}$:LiDFOB:40 wt. % POSS-PEG polymer electrolyte membranes. With permission from Ref. [51], Copyright © 2017 Elsevier.

This increase may be attributed to the increased free charge carriers and the increased segmental motion of polymer chains. The increase in temperature increases the polymer flexibility and hence the increased conductivity that was evidenced by the reduction of activation energy from 0.594 eV to 0.433 eV with 40 wt. % nanoparticle. The electrochemical stability window of the prepared system was up to 4.7 V and is in the desirable range. The thermal stability window of the prepared PNC was ~200 ºC (Figure 10 a). The stress-strain curve evidences the increase of stress from 0.076 to 0.099 MPa and elongation break from 472 to 516 %. This may be due to the key role of nanofiller as a crosslinking center (Figure 10 b). The electrochemical analysis of the cell (Li/PEO:LiDFOB:x wt% POSS-PEG(x = 0 and 40)/LCO-CBL) shows initial capacity up to 187 mAh/g (for 0 wt. % nanofiller is 158 mAh/g) and after 50 cycle discharge capacity was 143 mAh/g (for 0 wt. % nanofiller is 122 mAh/g) with coulombic efficiency of 99 % at 25$^{th}$ cycle..

Another report by Wang et al., [52] explored the role of $Li_{1.5}Al_{0.5}Ge_{1.5}(PO_4)_3$ (LAGP) on the PEO (LiTFSI) in the suppression of Li dendrite growth. XRD evidences the decrease of the crystallinity. Figure 11 a shows the process for the cell fabrication. Figure 11 b shows the ion transport mechanism on the addition of LAGP (for two system) and the preferred path is a path having low activation energy. The voltage stability window was broader (~5.12 V) as compared to pure PEO. The electrochemical performance was investigated for the cell composition Li-PEO (LiTFSI)/ LAGP-PEO1/LiMFP within the voltage 2.5−4.5 V (@0.2 C, At 50 ºC) as shown in Figure 11 c. The initial discharge capacity was 161.7 mAh $g^{-1}$ (Coulombic efficiency of 92.4%) and after 10 cycles the Coulombic efficiency was above 99%. Figure 11 d shows the rate discharge performance of the Li-PEO-500000- (LiTFSI)/LAGP-PEO1/LiMFP cell at 50 °C (@ 0.1 C, 0.2 C, 0.5 c & 1.0 C). Even at a high rate (1.0 C) the discharge capacity was 115 mAh $g^{-1}$. Figure 11 e show the CV, which confirms the reversible process of $Li^+$ extraction and insertion. The oxidation peaks at 3.6 and 4.23 V (attributed to $Fe_2^+$ to $Fe_3^+$ and $Mn_2^+$ to $Mn_3^+$) and the reduction peaks at 3.5 and 3.96 V ( Attributed to the reduction of $Fe_3^+$ to $Fe_2^+$ and $Mn_3^+$ to $Mn_2^+$) agrees well with the charge−discharge plateaus as shown in Figure 11 c.

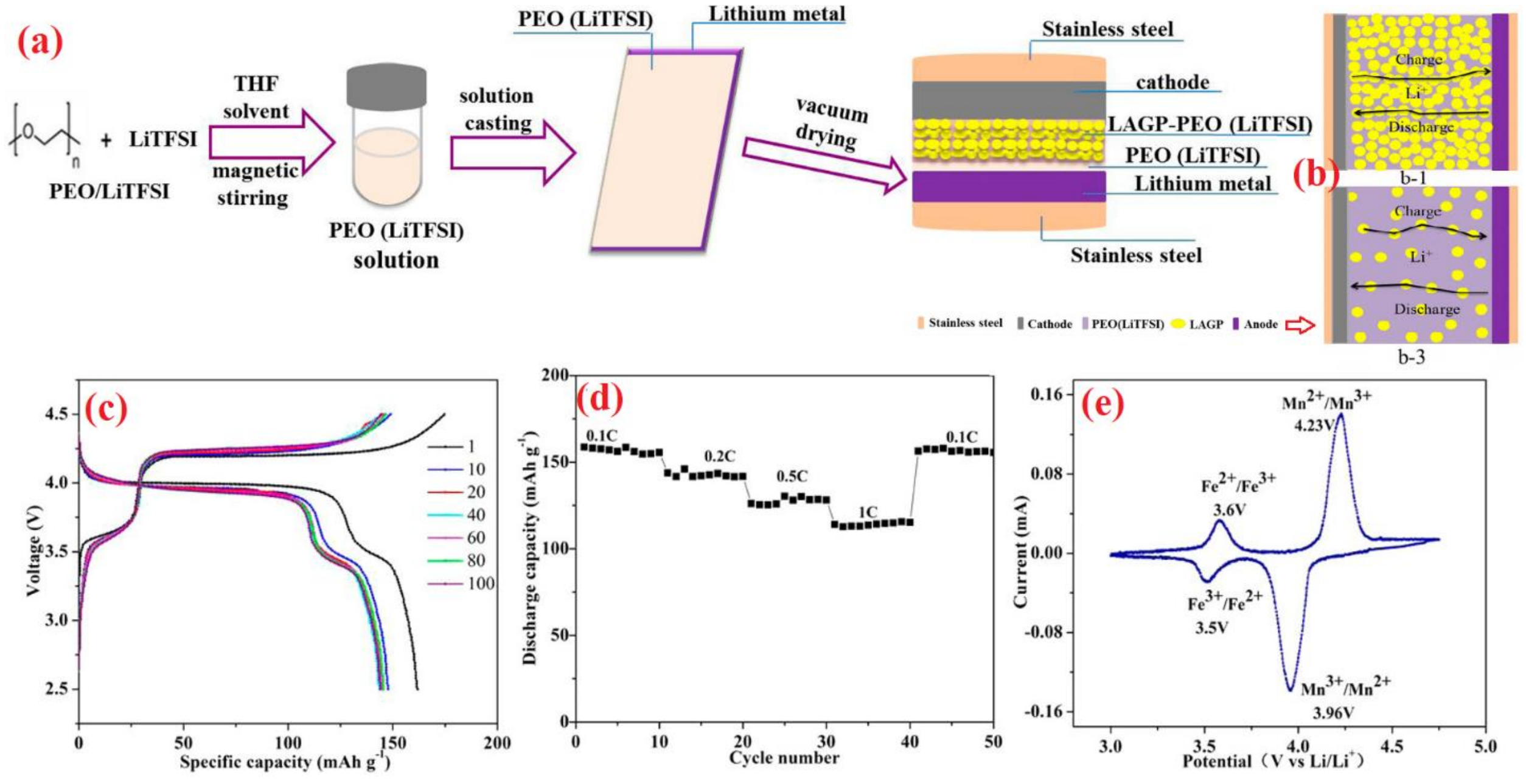

Figure 11. (a) All-solid-state Li-PEO (LiTFSI)/LAGP-PEO (LiTFSI)/LiMFP cells, (b) ) $Li^+$ ion transport mechanism in the composite solid electrolyte with different contents of PEO (LiTFSI): (b-1) LAGP-PEO1, (b-3) LAGP-PEO$^5$, (c) Charge/discharge curves. (d) Rate performance of Li-PEO-500000(LiTFSI)/LAGP-

PEO1/LiMFP cell (cutoff voltage: 2.5−4.5 V, 50 °C). (e) CV curve of LiMFP at a scan rate of 0.1 mV $s^{-1}$ in 2.5−4.75 V at 50 °C. With permission from Ref. [52] Copyright © 2017 American Chemical Society.

Arya et al., [53] investigated the effect of various nanofiller ($BaTiO_3$, $CeO_2$, $Er_2O_3$, $TiO_2$) on the PEO-PVC blend polymer electrolyte. The XRD analysis confirms the polymer nanocomposite formation. FTIR provides evidence of interaction among the functional groups of the polymer with the ions and the nanofiller in terms of shifting and change of the peak profile. The highest ionic conductivity is $2.3 \times 10^{-5}$ S $cm^{-1}$ with a wide electrochemical stability window of ~ 3.5 V for 10 wt. % $Er_2O_3$. Figure 12 shows the proposed ion transport mechanism.  It depicts that the anion is going to coordinate with the polymer backbone while cation with the ether group of the polymer chain. Nanofiller with the surface group also helps in the salt dissociation and polymer-ion-nanofiller interaction enhances the overall ion transport. Also, the coordinating interaction of the cation with the polymer chain modifies the polymer chain arrangement and disorder is produced that evidences the increase in the polymer chain flexibility. The enhanced flexibility is an indication of the enhanced conductivity, and the fast segmental motion of the polymer chain provides a path for ion transport.

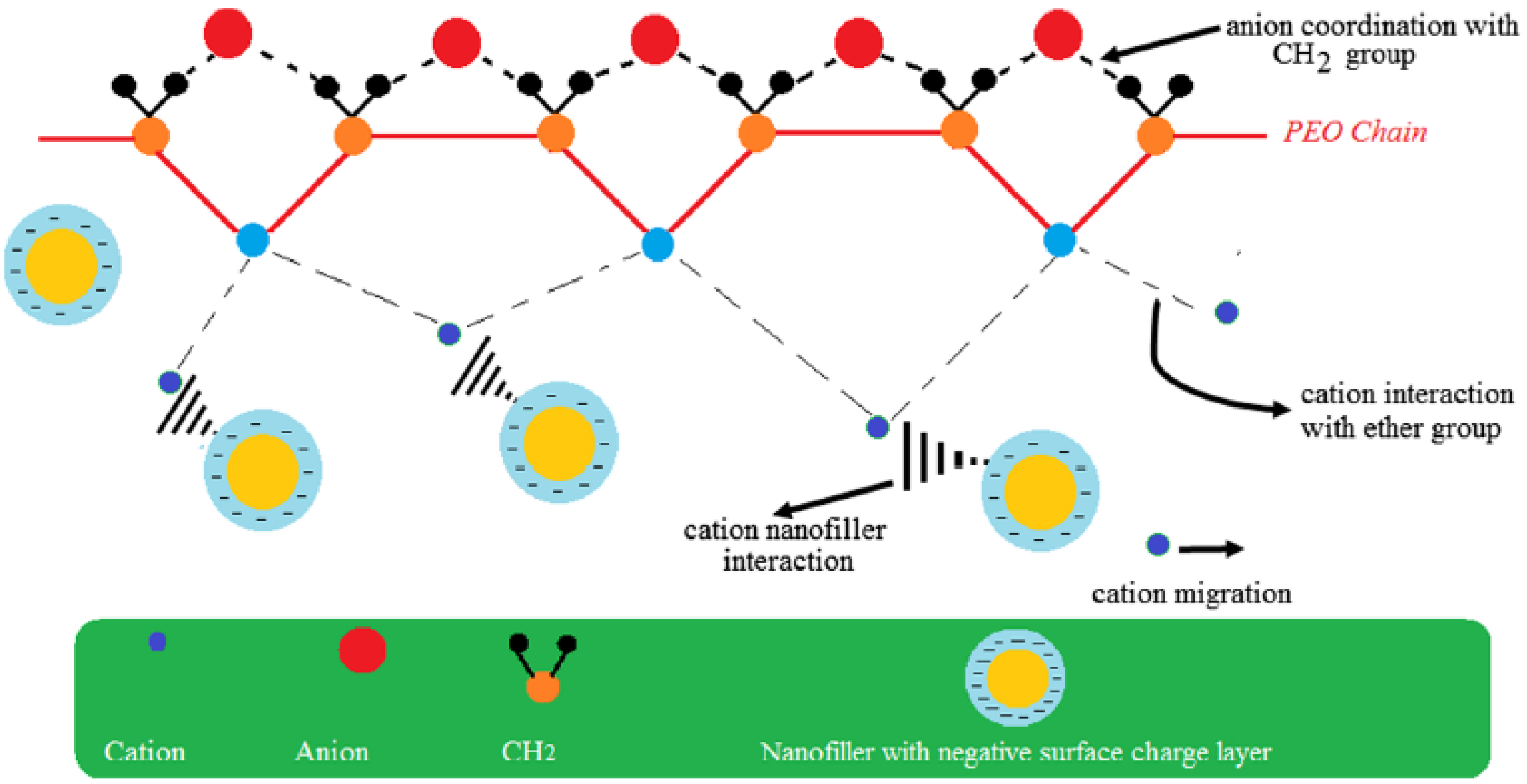

Figure 12. Proposed interaction scheme in the polymer nanocomposite matrix. With permission from Ref. [53] Copyright © 2017 Springer.

### 4.2. Nanoclay dispersed polymer nanocomposites

As studied earlier that the nanofiller enhances the possibility of interaction between polymer-ion via the Lewis-Acid-Base interactions. It alters the polymer chain arrangement and the more free volume is available for the ion transport. Another important role is played by the surface group of the nanofiller and it leads to the formation of the conductive continuous network. The overall enhancement is of the amorphous content that is beneficial for a fast solid state ionic conductor. But with nanofiller one issue is still there that is of dual ion conduction. So, an alternative is the use of nanoclay instead of the nanofiller.

One important step before using the clay is its modification in which the covalent bonding between the clay layers is disrupted by the introduction of the surfactants or hydrophobic functional moieties. This increases the dispersion of nanoclay and overall aim is to increase the basal spacing so that polymer chain can be intercalated easily (Figure 13 a). Two approaches are shown in Figure 13 b, (i) ion-dipole method, (ii) ion-exchange reaction [54-55].

Nanoclay has two advantages, one it blocks the anion migration inside clay galleries (the only cation is available for conduction) and another is that as polymer chain is intercalated inside the clay galleries so the polymer recrystallization tendency or crystallinity is reduced. Both the above properties lead to enhancement of the ion transport and hence the ionic conductivity. The high cation exchange capacity of the clay supports the intercalation and swelling of the polymer chains.

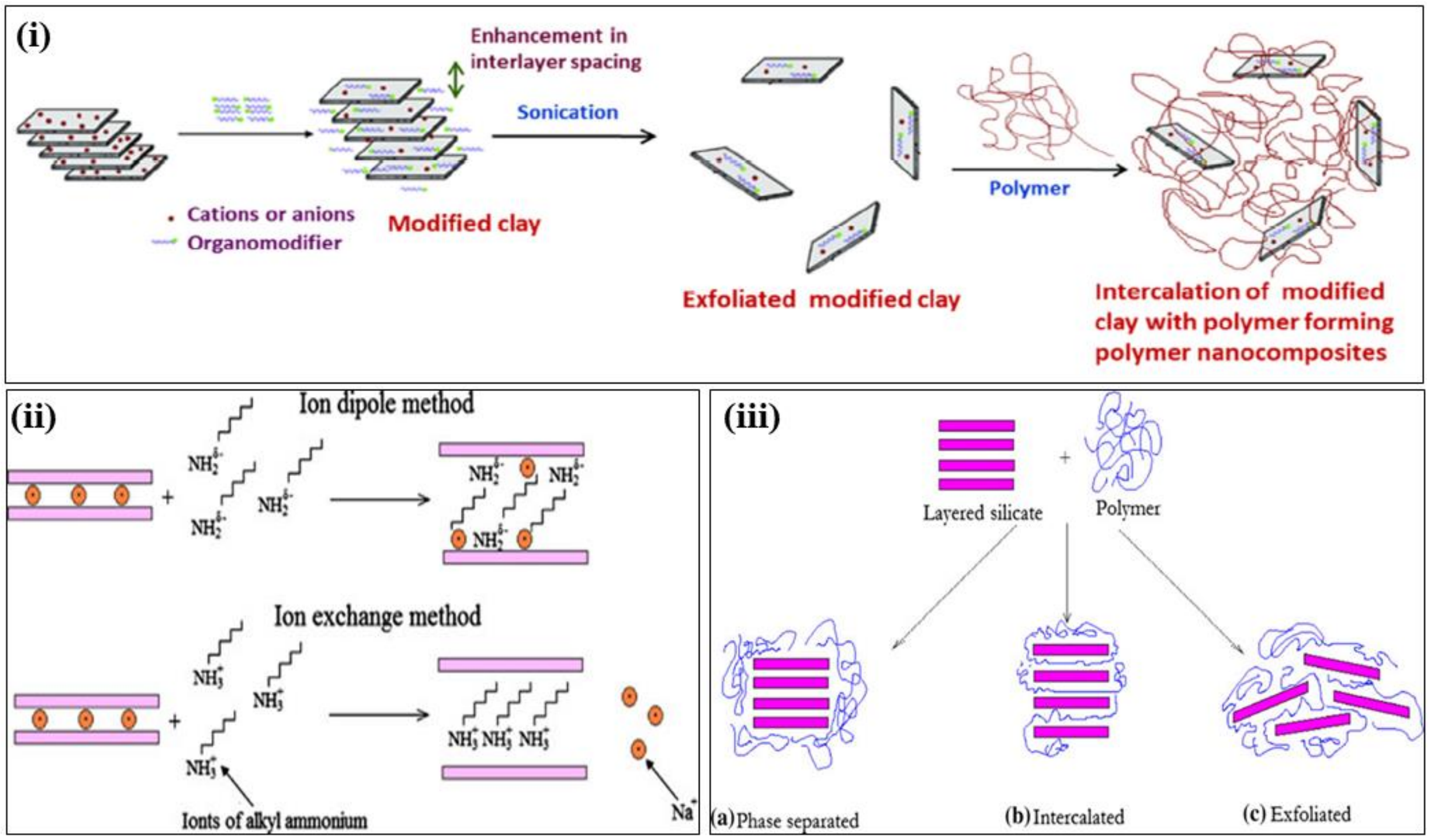

Figure 13. (i) Schematic diagram showing clay modification and intercalation of polymer to form polymer nanocomposites. With permission from Ref. [56] Copyright © 2015 Elsevier. (ii) Principles of modification of clay minerals and (iii) (The level of intercalation/exfoliation of nanofiller in a polymeric matrix. With permission from Ref. [55] Copyright © 2010 Springer.

The combined effect of them increases the dispersion of the nanoclay with the polymer matrix. By monitoring the separation of layers (basal spacing; $d_{001}$) four types of polymer nanocomposites with nanoclay are obtained (Figure 13 b) [56-58].

(i) If there is no change in the basal spacing with the addition of clay and clay layers remains outside then the PNC is *conventional* PNC.

(ii) If the basal spacing increases with nanoclay and clay layers are stacked with polymer chain intercalated between layers then it is called *intercalated* PNC. This type of the PNC leads to building up of the nano-metric channels for cation transport and disrupts the recrystallization tendency of the polymer chains.

(iii) If the clay layers are completely in the disordered state as well as the polymer chains then the PNC is *exfoliated* PNC. This type of PNC lowers the ion-pairing effect.

(iv) If the long molecular chains get intercalated inside the two or more clay galleries then it is called *flocculated* type PNC.

Beside this, the nanoclay such as halloysite nanotube (HN), montmorillonite (MMT) clay are gaining more intention in the formation of the polymer nanocomposites and main influence is on the thermal, electrical and the mechanical properties. The interaction mechanism behind the nanoclay is the formation of the electrostatic interactions between the charges present on the nanoclay surface and the electron rich group of the host polymer. This interaction disrupts the weak dipolar and van der Waals forces between clay sheets. As, now polymer get intercalated between the clay sheets and prevents the direct impact on the polymer on heating and hence the improved properties are achieved [59-65].

In the formation of the PNC preparation method plays an effective role in altering the chain arrangement. So, Dhatarwal et al., [66] reported the preparation of the polymer nanocomposites (PNC) solid polymer electrolyte based on PEO-PMMA-$LiBF_4$+ 10 wt. % EC and 3 wt. % MMT clay by solution-cast (SC) and the ultrasonic-microwave irradiated (US–MW) solution-cast methods. XRD analysis suggests the complete dissolution of the salt as there was no peak corresponding to the salt (2θ = 26.33°) (Figure 14). Pure MMT shows the peak at 2θ = 7.03° (001; plane) and is also observed in the PNC prepare by SC technique along with two main peaks of PEO (2θ =19.43° and 2θ = 23.57°,).

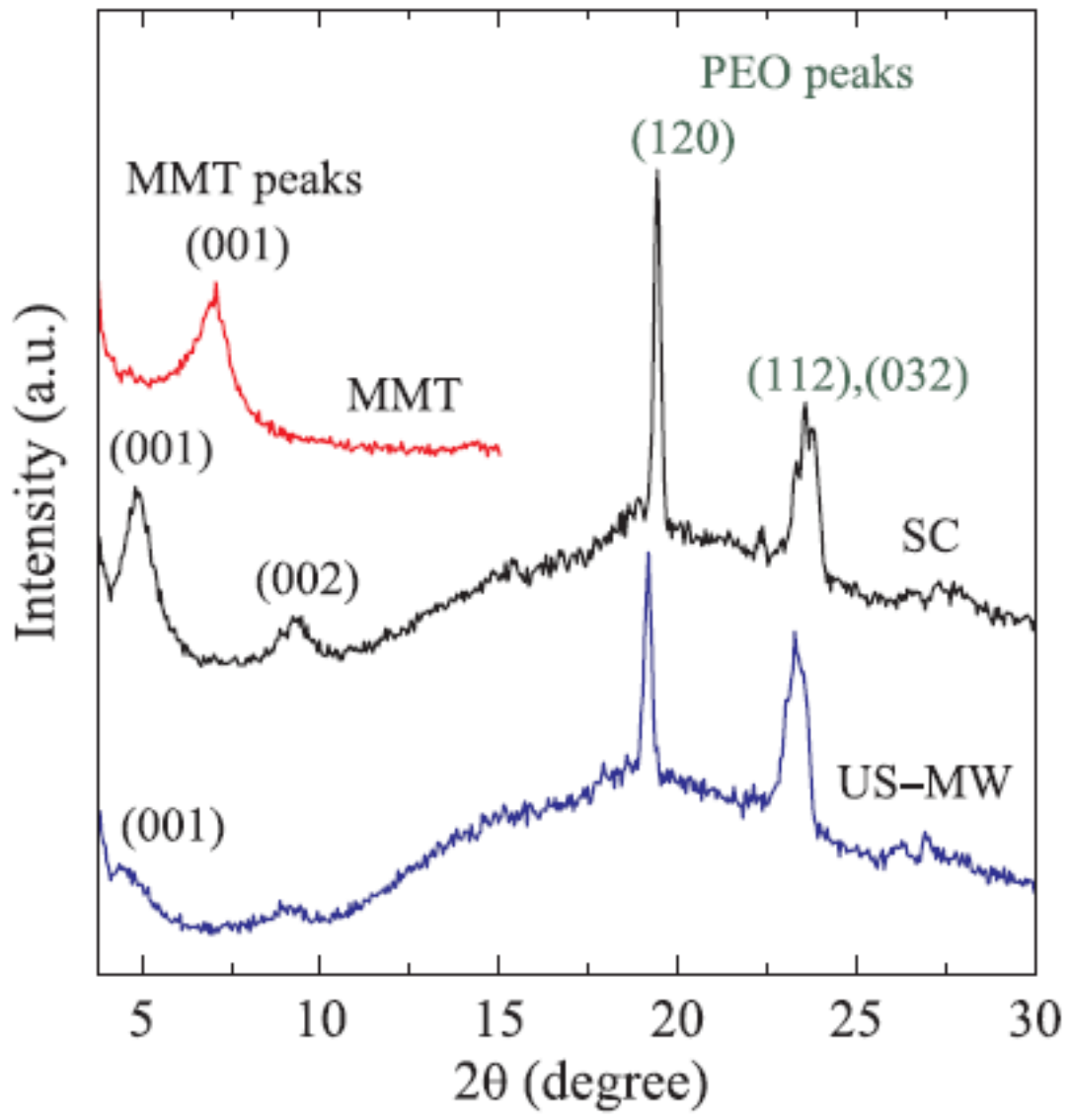

Figure 14. XRD patterns of MMT nanopowder and (PEO–PMMA)–LiBF4–10 wt. % EC–3 wt. % MMT films prepared by SC and US–MW methods. With permission from Ref. [66] Copyright © 2017 Elsevier.

Further for the PNC, the MMT peak shifts toward lower angle by more than 2° as compared to that of the pristine MMT and increase of the d-spacing is evidence the formation of the intercalated structure. The SC methods display the high intensity of MMT while US-MW methods show a diffused peak with low intensity. The former one is evidence of the intercalated structure while later one is an indication of the exfoliated structures [67]. It was concluded that the increased basal spacing, crystallite size, and the relative intensities were increased for the PNC prepared by the US-MW method. Further, from the impedance spectra, it was founded that the conductivity was more for the intercalated type PNC and supports faster ion migration as compared to the exfoliated type PNC and hopping mechanism seems to be followed as indicated by the value of n. So, the preparation methods influence the ion mobility and intercalated type is more beneficial as compared to the exfoliated type which hinders the ion migration.

Erceg et al., [32] prepared the composite polymer electrolyte based on poly(ethylene oxide)/lithium montmorillonite (PEO/LiMMT) by melt intercalation technique. The small angle X-ray scattering (SAXS) evidenced the increase of the interlayer spacing due to polymer chain intercalation inside the clay galleries with a maximum of about 1.88 nm while the interlayer distance was 0.93 nm (maximum). DSC analysis displays the powering of the melting temperature and indicates the disruption of the crystallinity after addition of clay (76.1 % to 37.1 %). FTIR spectrum evidences the broadening in the spectrum in region 3000 and 2750 $cm^{-1}$, and 1500 $cm^{-1}$ to 2000 $cm^{-1}$ while some new peaks were observed after clay addition. It confirms the existence of a crystalline phase and gets broadened with clay addition indicating the change of the crystallinity. The highest ionic conductivity was $2.8\times10^{-6}$ S cm-1 for 40 wt. % clay content and may be due to the proper dispersion of nanoclay while at higher content self-aggregation of clay layers trap the cation and hence the lowering of the ion mobility.

Sengwa et al., [68] prepared the solid polymer nanocomposite electrolytes (SPNEs) based on poly(methyl methacrylate) (PMMA) and lithium perchlorate ($LiClO_4$) with varying concentration of montmorillonite (MMT) clay by solution casting and high intensity ultrasonic assisted solution casting methods. XRD analysis evidences the complete dissociation of the salt and the exfoliation of clay was attributed to the interactions of polymer salt complex (C=O---$Li^+$) with the MMT Nanosheets surfaces. Impedance analysis suggested that the current carriers are ions which govern the total electrical conductivity of these electrolytes with value about $10^{-5}$ S $cm^{-1}$. Figure 15 a displays the impedance of the sample prepared by the solution cast method. The proposed mechanism supports the enhancement of the conductivity with the addition of clay. Figure15 b shows the cation coordination with the C=O in PMMA–$LiClO_4$ domain while anion is somewhere in the polymer backbone as an un-coordinated form. Figure 15 c displays the cation coordination with the MMT domain and exfoliated structure is observed. Figure 15 d demonstrates the PMMA----$Li^+$----MMT interactions.

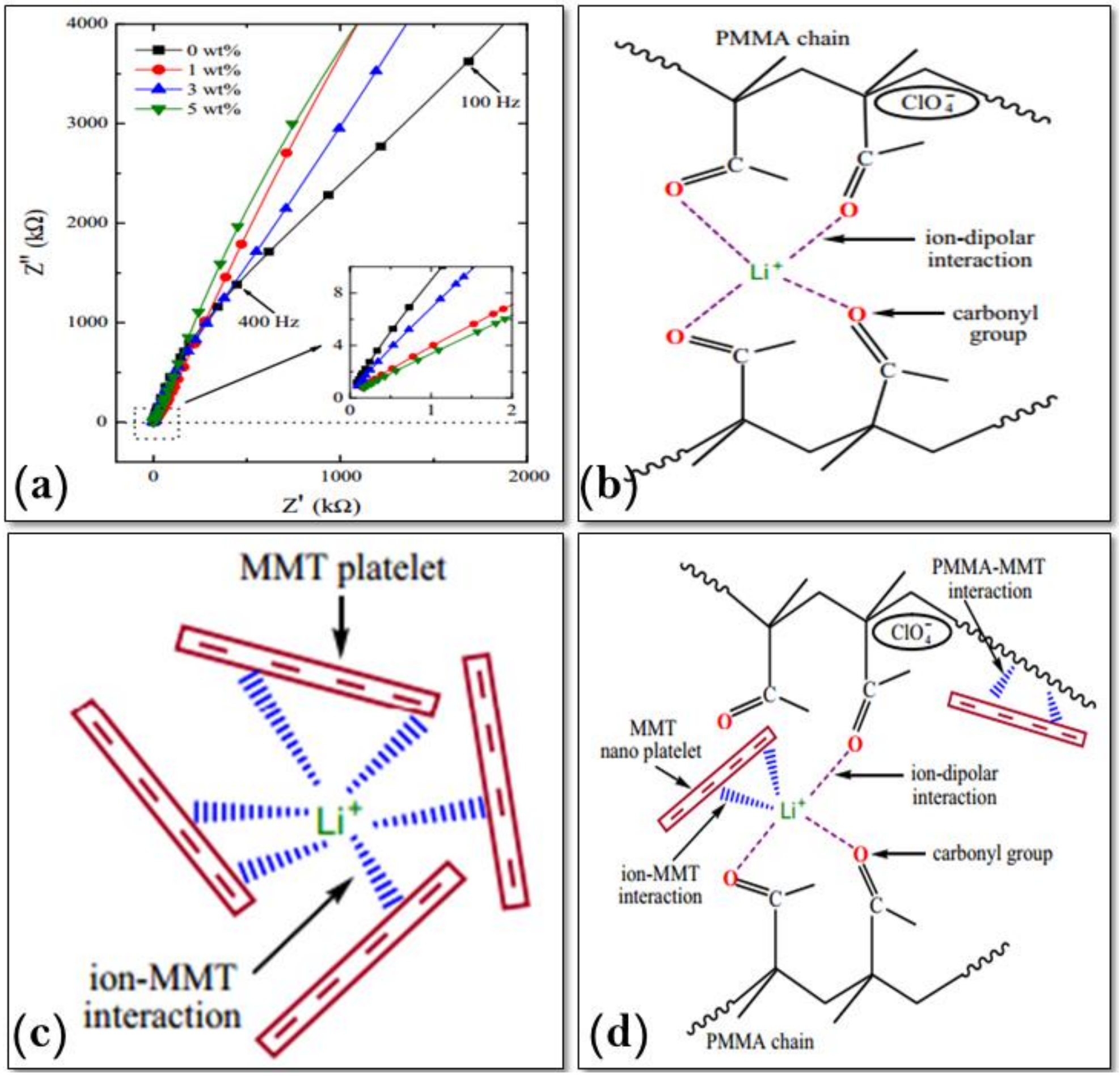

Figure 15. (a) Complex impedance plane plots (Z″ vs. Z’ ) of PMMA-$LiClO_4$–x wt. % MMT films prepared by solution cast (SC) method and Schematic illustration of PMMA–LiClO4–MMT interactions: (b) C=O---$Li^+$ complexes, (c) $Li^+$---MMT complexes and (d) C=O---$Li^+$ ---MMT complexes. With permission from Ref. [68] Copyright © 2014 Elsevier.

Lin et al., [69] reported the preparation of nanocomposite using halloysite nanotube (HNT; 10–50 nm outer diameter and 5–20 nm inner diameter, with a length of 50–1000 nm) clay in PEO-LiTFSI based polymer matrix. The highest ionic conductivity was $1.11\times10^{-4}$ S $cm^{-1}$, $1.34\times10^{-3}$ S $cm^{-1}$, $2.14\times10^{-3}$ S $cm^{-1}$ at 25, 60 and 100 °C respectively for 10 % HNT (Figure 16 b). This increase of the conductivity was evidences further by the optical images which show the decrease of crystal size with the addition of HNT ($Al_2Si_2O_5(OH)_4$). The $Li^+$ transference number was also increased from 0.25 to 0.40 on the addition of HNT. The voltage stability window of the composite polymer electrolyte was 6.35 at 25 $^oC$ and decreased to 4.78 V at 100 $^oC$ (Figure 16 c). Figure 16 b explores the increase of conductivity with the addition of HNT. The HNT has two face surfaces; one has a -Si-O-Si- silica tetrahedral sheet and another has -Al-OH groups from the octahedral sheet known as outer and inner surface respectively (Figure 16 a). The presence of opposite charge on the HNT surface separates the ion pairs and cation get associated with the negatively charges silica surface (-Si-O-Si-) while the anion get associated with the inner surface (-Al-OH). Also, the electron-rich ether group of PEO interacts with the cation and 3 D network formed here supports the ion migration. The overall impact is the disruption of the crystalline phase due to the reduction of the crystallinity and hence the low ion pair formation. One another point mentioned here was that HNT helps in improving the mechanical property with uniform surface. Also, the Zeta potential measurements show increase from negative to positive and this evidences that adsorption of ions on the HNT surface. The thermal stability window was ~400 °C. The stress after addition of HNT increase from 1.25 MPa to 2.28 MPa with 400 % strain and displays good flexibility of the investigated polymer matrix. The initial discharge capacity of the cell was 1350 mAh $g^{-1}$ and with an average value of 745 ± 21 mAh $g^{-1}$ in the 100 discharge/charge cycles, with 87% retention. While at 100 °C the initial discharge capacity was 1493 mAh $g^{-1}$ and was 386 mAh $g^{-1}$ after the 400 discharge/charge cycles. It was concluded here that the natural HNT clay mineral provides the desirable energy density at low cost with improved safety.

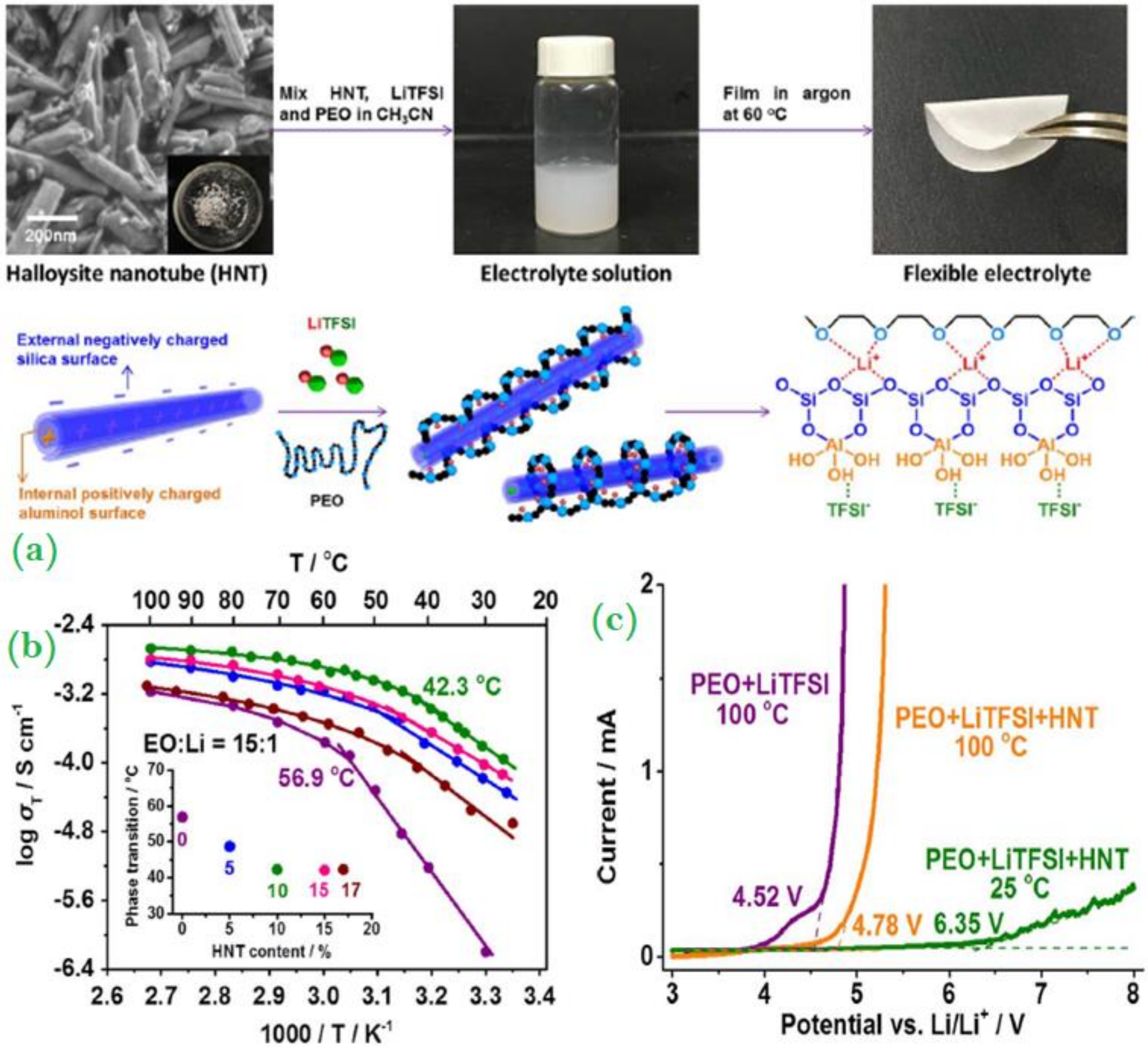

Figure 16. (a) Preparation of HNT modified flexible electrolyte and mechanism of HNT addition for enhanced ionic conductivity. The halloysite nanotube, LiTFSI and PEO are mixed in the solvent to form a uniform electrolyte solution. The solution is cast in an argon atmosphere to produce a flexible electrolyte thin film, and Lithium ion transport for HNT nanocomposite electrolytes. (b) Ionic conductivities of the PEO+LiTFSI+HNT films with different HNT

contents at EO:Li=15:1 as a function of temperature, inset is the phase transition temperature as a function of HNT content obtained after fitting. (c Linear sweep voltammetry Li/PEO+LiTFSI+HNT/SS cells at 25 °C and 100 °C, and Li/PEO+LiTFSI/SS cells at 100 °C at a rate of 10 mV $s^{-1}$. With permission from Ref. [69] Copyright © 2017 Elsevier.

Gomari et al., [70] reported the preparation of nanocomposite solid polymer electrolyte based on poly(ethylene) oxide (PEO) and lithium perchlorate salt ($LiClO_4$) with pristine graphene (GnP) or polyethylene glycol-grafted graphene (FGnP). From FESEM analysis of the nanocomposite, it was concluded that the rough surface of the pristine PEO changes to smooth on the addition of GnP/FGnP and suggests the better dispersion of both. Another reason behind this may be the presence of the hydrogen interactions between the electron rich group of host polymer and PEG groups of FGnP. XRD analysis evidences the decrease of the crystallinity with the addition of GnP and more decrease was observed for the FGnP based polymer nanocomposite. Further evidence provided by the DSC analysis demonstrates no change in the glass transition/melting temperature with a dispersion of GnP addition while FGnP dispersion displays shift toward lower temperature of both glass transition/melting temperature. This may be associated with the better dispersion of the FGnP as compared to the GnP which enables sufficient interactions with the PEO. The lowering of the $T_g$ indicates an increase of ion mobility or polymer flexibility owing to the increased free volume. The Polarized optical microscopy also supports the both XRD and DSC results as suppression of spherulitic growth in the presence of graphene nanosheets were obtained as compared to PEO. Further deconvolution of the anion peak in the FTIR spectra confirmed the increase of free charge carriers with FGnP which participate in conduction and was attributed to the interaction of PEO as well as additional coordinating sites provided by PEG. The ionic conductivity was $8.19\times10^{-6}$ S $cm^{-1}$ for 0.1 % GnP and $2.53\times10^{-5}$ S $cm^{-1}$ for 0.5 % FGnP. The increase in the conductivity was associated with the additional coordinating sites provided by the PEG and an ion conduction channel formation supports the faster segmental motion. The stress-strain curve reports the increase of stiffness and was 74% and 83% increase for GnP and FGnP at 0.5 wt %, respectively.

Kim et al., [71] reported the preparation of the polymer composite electrolyte based on the poly(ethylene oxide) (PEO)-$LiClO_4$ and different content of the lithium montmorillonite (Li-MMT) clay. The absence of the Na peak in EDS spectra evidences the de-intercalation of Na and the intercalation of the Li-ion. XRD analysis suggested the decrease of the interlayer spacing from 32.43 to 25.66 Å with increase of the clay content from 10 to 25 wt.% and evidences the difficult for polymer intercalation in the clay gallery. Further for 20 wt. % of nanoclay content fundamental peak of PEO is broadened and intensity is reduced. It, evidences the decrease of the crystallinity value. This was further supported by the DSC data which demonstrates the shift of the melting peak toward lower temperature (for pure PEO 65.4 °C to 52.2 °C) and decrease of crystallinity with lowest for 20 wt. % clay content (for pure PEO 53.97 % to 37.73 %). Figure 17 a depicts the DSC results with different clay content. The thermal stability of the PCE was up to 200 °C. The highest ionic conductivity was obtained for the 20 wt. % clay content and is about $5.3\times10^{-6}$ S $cm^{-1}$. Further, the temperature variation of the conductivity follows the Arrhenius behavior and activation energy decreases with the addition of clay content (0.70 to 0.19 eV, below 60 °C). The cation transference number ($t_{Li+}$) was 0.55 and this high value of cation transference number was attributed to the poor interaction of cation with ether group of PEO owing to the clay interaction with ether group.

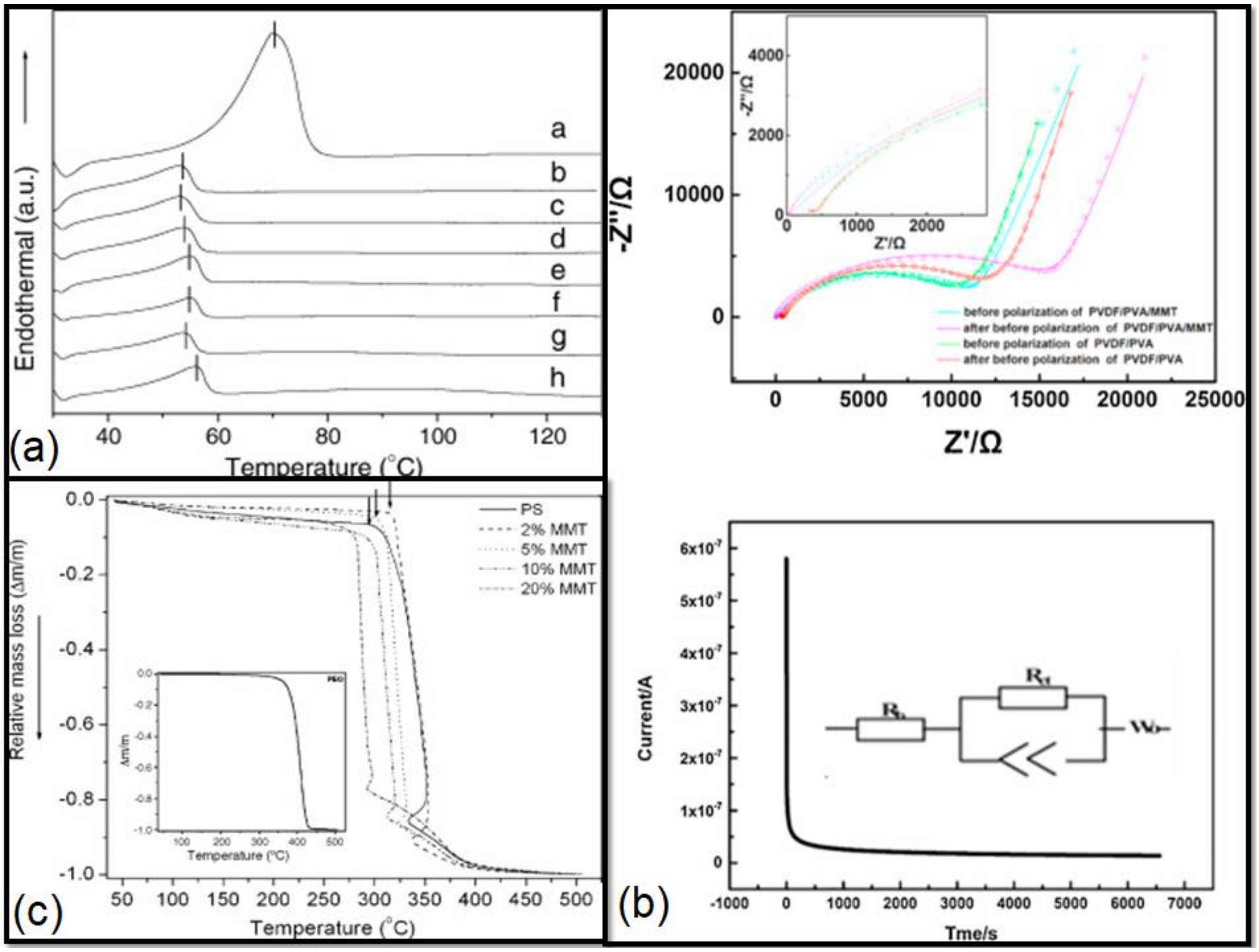

Figure 17. (a) DSC curves of (a) pure PEO and PCEs containing (b) 0%, (c) 2%, (d) 5%, (e) 10%, (f) 15%, (g) 20%, and (h) 25% of Li-MMT. With permission from Ref. [71] Copyright © 2007 Elsevier., (b) AC impedance behaviors of polymer electrolytes based on casting PVDF/ PVA/MMT CSPE. With permission from Ref. [72] Copyright © 2016 Elsevier.and (c) Thermogravimetry analysis (TGA) plots of polymer–salt complex (PS) and polymer based nanocomposite (PNCE) films having different clay concentration compared with TGA pattern of pure PEO (inset). With permission from Ref. [73] Copyright © 2009 Elsevier.

Ma et al., [72] prepared the composite solid polymer electrolyte (CSPE) based on montmorillonite (MMT) nano-clay fillers, lithium-bis(trifluoromethanelsulfonyl) (LiTFSI), polyvinylidenedifluoride (PVDF) and polyvinyl alcohol (PVA) copolymer by the casting method. FESEM analysis evidences the porous structure that will be beneficial for the cation transport. Further, the addition of MMT to blend polymer matrix lowers the crystallinity and indicates the increased amorphous content. The ionic conductivity also increases with the addition of MMT and highest conductivity was $4.31\times10^{-4}$ S/cm for 4 wt. % clay content. The increase in the conductivity value was due to the high surface area of the nanoclay which increase in the free volume and the amorphous content. These together leads to smoother ion transport and hence the high ionic conductivity. The activation energy value decreases with the addition of MMT from 26.46 kJ to 16.22 kJ and evidences the faster ion migration. The cation transference number increased from 0.29 to 0.40 with the addition of the MMT (Figure 17 b). The displacement load curve evidenced the increase of the tensile strength from 1.17 MPa to 2.24 MPa with the addition of clay and is attributed to the improved the stability. The Li/CSPE/LiFePO4 cells show pretty high specific discharge capacity above 123 mAh/$g^{-1}$ along with a coulombic efficiency of 97.1% after 100 cycles.

Mohapatra et al., [73] reported the preparation of the PEO-$LiClO_4$ based composite polymer electrolyte with organo-modified montmorillonite clay using solution cast technique. XRD analysis evidenced the complex formation and shift in the peak of the hots matrix shows the effective role played by clay. The decrease of the peak intensity evidences the reduction of the crystallinity due to the polymer and the nanoclay interaction. The d-spacing was almost same in all the systems while the crystallite size decreases and clay gallery width was maximum for the 10 wt. % clay content (9 Å). Further, the polymer salt intercalation in the clay galleries was evidenced and was due to the dipolar interaction. It was concluded from the XRD that the prepared PNCE films have a multiphase combination of crystalline and amorphous PS phases, an amorphous phase boundary, crystalline clay and amorphous PS inside the clay galleries at the interface of the clay layers and PS matrix. Further TEM analysis provides strong evidence of the intercalation along with exfoliation at low clay content. At very high clay content (>20 wt. %) clay cluster formation was reported. DSC analysis displayed the lowering of the glass transition temperature ($T_g$) owing to the enhanced flexibility assisted by intercalation on nanocomposite formation. The highest ionic conductivity was $6.48\times10^{-5}$ S/cm for 10 wt. % clay content (at 30 $^{o}$C). TGA analysis shows the thermal stability of the composite polymer electrolyte up to 300 oC and is sufficient for the application purpose (Figure 17 c). The cation transference number was 0.50 at 2 wt. % clay content. And voltage stability window was about 3 V. The thermal stability was improved after the polymer chain intercalation inside clay galleries. This may be attributed to the barrier role played by clay layers which prevent the decomposition of the polymer salt complex.  Another reason may be the catalytic effect in which the clay layers accumulated all heat and prevents the decomposition of the polymer slats system.

Sharma et al., [74] reported the preparation of the polymer nanocomposites based on $(PAN)_8LiCF_3SO_3$ -x wt %DMMT. From XRD analysis the increase in the interlayer spacing and the clay gallery width was observed. This increase was attributed to the polymer chain intercalation in the clay nanometric clay galleries.  The ionic conductivity was highest for the 7.5 wt. % clay content and is about $6.8\times10^{-3}$ S $cm^{-1}$. The ion transference number was close to unity and evidence the dominance of the ionic nature of the electrolyte. The cation transference number was 0.67 and is an indication of proper dissociation of the salt after polymer chain intercalation inside clay galleries. The voltage stability window was 5.6 V and thermal stability window was approx. 200 $^{o}$C.

Moreno et al., [75] prepared the polymer nanocomposite based on the PEO (with molecular weight 600,000 and 4,000,000)-bentonite. XRD analysis demonstrated the absence of peak associated with the bentonite confirmed the polymer intercalation inside clay galleries. Figure 18 shows the tensile stress-strain curves for PEO Mw 600,000 SPE films. The Young's modulus and the tensile strength (MPa) of the PEO/PEO@bentonite-$Li^+$ were 325± 32.5 MPa and 7.5 ± 0.75 MPa, while the PEO/bentonite-$Li^+$ was 247 ± 24.7 MPa and 2.9 ± 0.29 MPa. Compared with the PEO/bentonite-$Li^+$, the ionic conductivity of the PEO/PEO@bentonite-$Li^+$ increased from $3.89 \times 10^{-8}$ to $1.81 \times 10^{-7}$ S $cm^{-1}$ at 25 $^{o}$C. It can be seen that the mechanical properties improved after addition of the 3 wt. % bentonite.

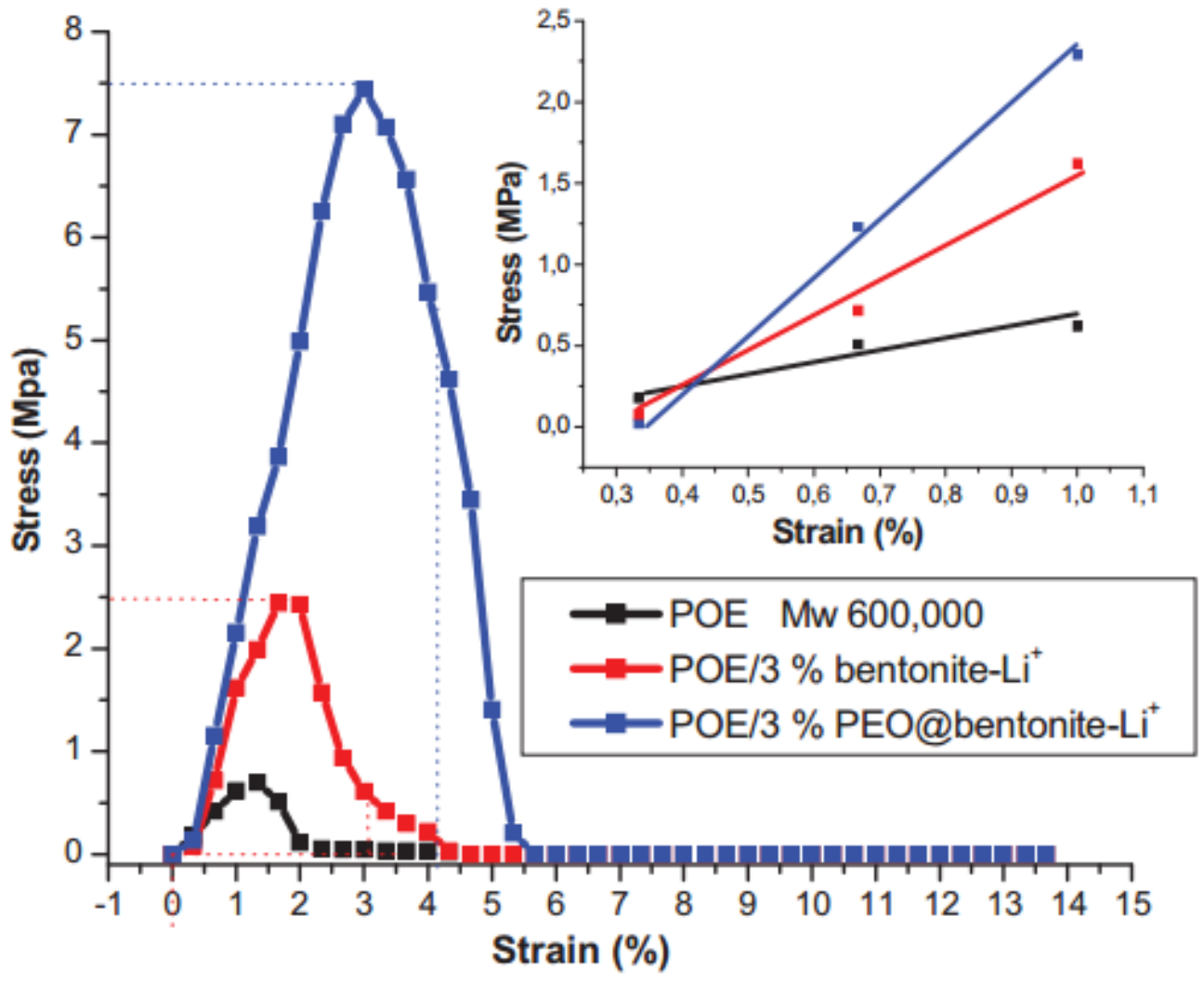

Figure 18. Tensile stress-strain behavior of films of PEO and SPEs with bentonite-Li+ and PEO@bentonite-$Li^+$ with PEO Mw 600,000. With permission from Ref. [75] Copyright © 2011 Elsevier.

Zhang et al., [76] reported the preparation of solid polymer electrolyte comprising of PEO, LiTFSI, and MMT. The XRD analysis evidences the increase of the interlayer spacing that confirms the polymer intercalation and hence increases of the amorphous content. The highest ionic conductivity was $3.22\times 10^{-4}$ S $cm^{-1}$ at 60 °C for 10 wt. % MMT and cation transport number ($t_{Li^+}$ ) was 0.45.

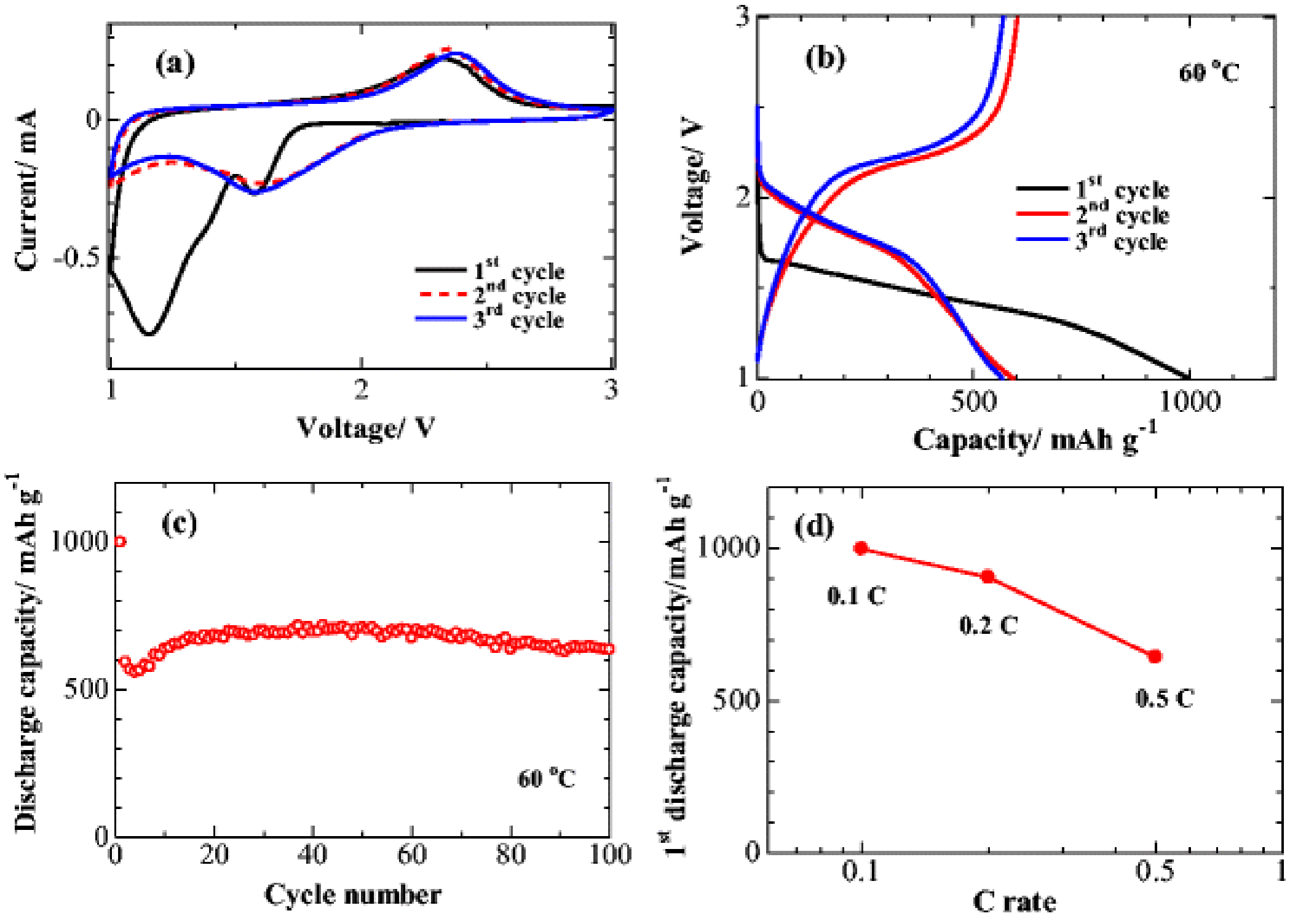

Figure 19. **a** Initial CV profiles of all solid-state Li/S cell at 60 °C; the measurement is conducted at a scan rate of 0.1 mV s−1 in the voltage range of 1.0 to 3.0 V vs. Li+/Li; **b** Charge/discharge profiles (at 0.1 C) of all solid-state

Li/S cell at 60 °C; **c** Cycle performance (at 0.1 C) of all-solid-state Li/S cell at 60 °C; **d** Rate capability of all-solid-state Li/S cell at 60 °C. With permission from Ref. [76] Copyright © 2014 Springer nature.

The voltage stability window of the optimized electrolyte was 4 V. The cyclic voltammograms (CV) of the cell shows good electrochemical stability in the operating range (Figure 19 a). The galvanostatic charge/discharge cycling tests in Figure 19 b, shows the specific capacity of 998 mAh $g^{-1}$ (@ 1st discharge at 0.1 C) and a reversible capacity of 591 mAh $g^{-1}$ (For 2nd cycle). Figure 19 c shows the initial increase in capacity which indicates the gradual activation of the electrochemical properties of polymer electrolyte electrochemical and is attributed to the cation dynamics in the polymer matrix. Even, after 100 cycles a high reversible specific discharge capacity (634 mAh $g^{-1}$) with 63.5 % capacity retention is noticed. Figure 19 d displays the rate capability of cells (@ 0.1, 0.2, and 0.5 C) and is sufficient for the fabricated cell.

### 4.3. Nanorod/Nanowire dispersed polymer nanocomposites

As a lot of reports are available on the dispersion of nanofiller in the polymer electrolyte matrix which provides in hand the superior properties as compared to the micro fillers. There are three fundamental mechanisms which dominate here, (i) creation of percolation pathways, (ii) reduction of polymer reorganization tendency and (iii) aspect ratio or shape of the nanoparticle. As nanofiller surface have acidic and basic sites which play an effective role in the ion transport or mobility in case of polymer electrolytes. These sites support the formation of conducting pathways for the ion transport. These paths are termed as percolation pathways. As the nanofiller surface groups alter the existing interaction between the polymer chains and the polymer ion. This lowers the chain reorganization tendency and provides smoother in transport. Another key role is played by the aspect ratio of the nanoparticle. As it is well known that the nanofiller have a high aspect ratio as compared to the micro fillers and the aspect ratio is in inverse relation with the percolation threshold. Also, the optimum or critical concentration is lower for the nanoparticles with high aspect ratio due to increased size distribution [77-81].

The mechanism behind the improved conductivity on the addition of nanofiller highlights the interaction electron rich group of host polymer with the surface group of nanofiller. This supports the salt dissociation and more free charge carriers are available for conduction. Another important role played by nanofiller is the disruption of the recrystallization tendency of the polymer chain arrangement and hence the amorphous content is increase which favors the smoother and faster ion migration. So, the percolation pathways created by the nanofiller are beneficial for the faster ion transport. It can be concluded from here that the larger the percolation path, faster will be the ion migration due to the long continuous path. This will depend on the coordinating sites (acidic/basic) available to favor the ion mobility and sufficient salt dissociation. Another remarkable point here is that the suitable control over the acidic/basic sites of nanofiller can lead to the improved electrical properties. Also, the alignment of the nanorod and nanowire is important as aligned nanoparticle parallel to the electrodes limit the perpendicular ion migration. The nanoparticles with the different aspect ratio and shape also alter the mechanical properties that are further altered by the alignment of nanorod/nanowire/nanotube.

Beyond the active/passive nanoparticle, one another attractive approach is the use of 1 D nanofiller for enhancing the electrical properties. So, regarding this Liu et al., [78] reported the dispersion of LLTO ($Li_{0.33}La_{0.557}TiO_3$) nanowire in the PAN-$LiClO_4$ based polymer salt matrix. XRD analysis provides no alteration in the peak of pure PAN with the addition of nanowire and SEM micrograph displays the uniform distribution of NW and are fully embedded in the polymer matrix evidenced by TEM (Figure 20 a-f). The highest ionic conductivity was $2.4 \times 10^{-4}$ S $cm^{-1}$ (15 wt. % LLTO) and is three orders higher than the polymer salt matrix without NW. AT high NW content lowering of conductivity is observed and may be associated with the aggregation of NW and incomplete dissolution of polymer with NW. Further, the dehydration temperature was correlated with the crystallinity and it was concluded that the disruption of the crystallinity was not the dominant factor for enhancement of the conductivity and is also confirmed by the XRD. This enhancement in the conductivity was explained with the oxygen vacancy-rich surface of the nanowire. These vacancies help in the salt dissociation and oxygen vacancy helps in smoother ion migration and hence the faster ion mobility. The voltage stability window was increased from 4.8 V (for NW free) to 5.5 V (wt. 10 wt. % NW).

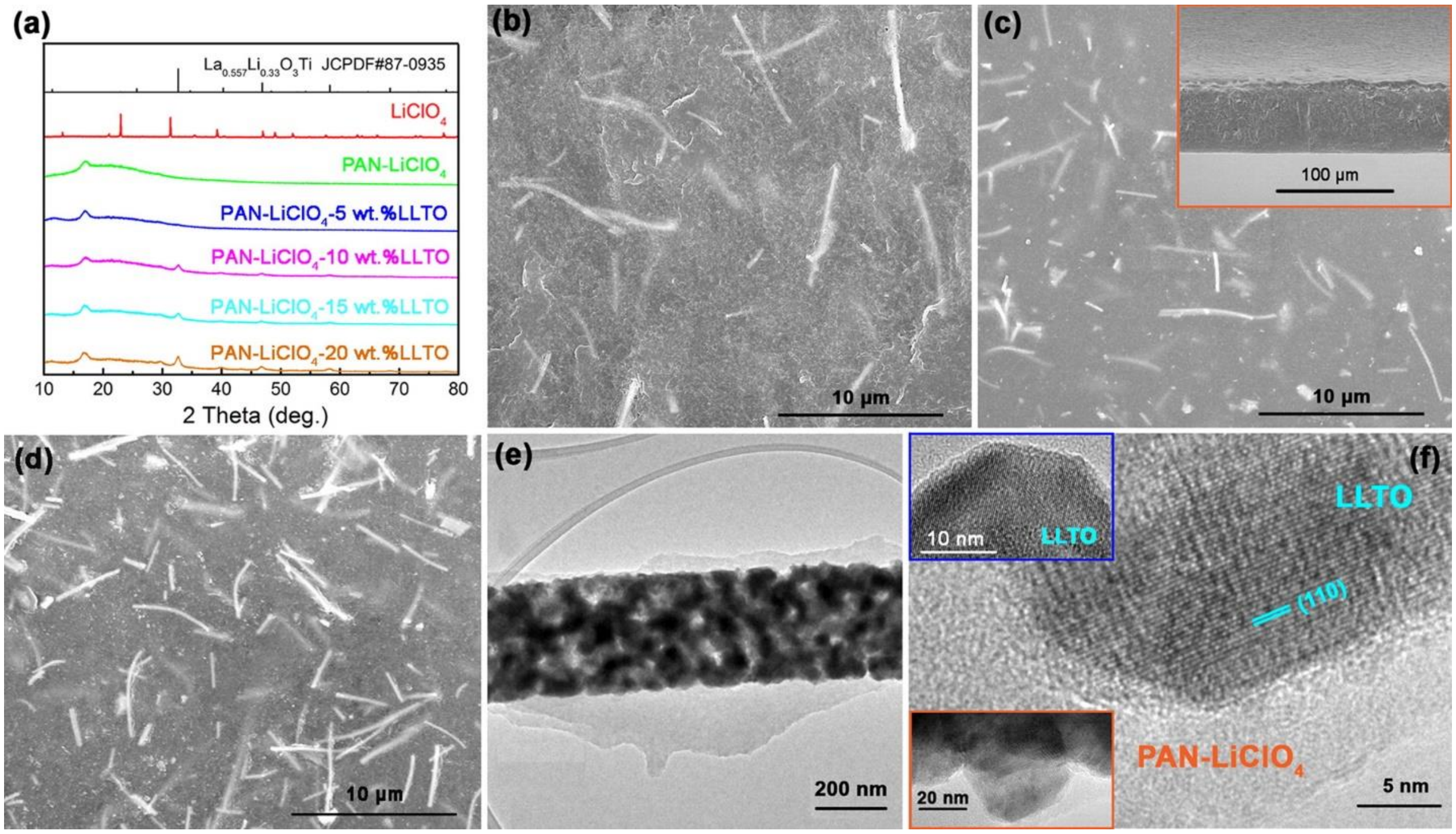

Figure 20. Phase structure and morphology of the composite electrolytes with various contents of LLTO nanowires. (a) XRD patterns of the composite electrolytes with various LLTO concentrations of 5−20 wt %. SEM images for the composite electrolytes with (b) 10 wt %, (c) 15 wt %, and (d) 20 wt % nanowire fillers. (e) TEM image and (f) HRTEM image of the composite electrolyte with 15 wt % nanowires, respectively. In panel f, the upper inset is the HRTEM image for LLTO nanowire, and the bottom one illustrates the individual grain of the nanowires embedded in PAN matrix. With permission from Ref. [78] Copyright © 2015, American Chemical Society.

So, keeping this in the mind Do et al., [79] prepared the nanorod ($\alpha$-$Fe_2O_3$ nanofiller with aspect ratio 7),) so that the desired improvement can be achieved at low critical concentration. For better comparison both nanoparticle (NP; Diameter 20-30 nm) and nanorod (NR; Diameter 10-20 nm) were dispersed separately. As nanorod has a higher aspect ratio which is associated with the longer percolation paths available for the conduction. Figure 21 a & b depicts the FESEM image of NR (Average length and diameter of 105 ± 32 and 16 ± 5.4 nm, Aspect ratio 6.6) and NP with average diameter 29 ± 11 nm. The ionic conductivity variation with the temperature suggests that the maximum conductivity was achieved with NR even at a lower content than the NP (Figure 21 c & d). This may be attributed to the formation of a sufficient number of conducting pathways at low concentration owing to the high aspect ratio of NR as compared to NP. Also, at high concentration, there is a decrease of conductivity and that may be due to the formation of ion traps because of rods aligned in parallel with the electrodes. Above the melting temperature, the enhancement in the conductivity was obtained. From, DSC results it was concluded that the nanofiller does not affect the polymer chain flexibility as no change in glass transition temperature ($T_g$) was evidenced. This was explained using the activation energy concept and is due to suppression of the crystallinity as evaluated form the melting peak. It was concluded that the recrystallization tendency is more for the NR (1 wt. %) due to more inter-particle interaction as compared to the NP. This leads to more aggregation of NR as compared to NP. So, here it was explained on the basis of the length of conducting pathways of NR.

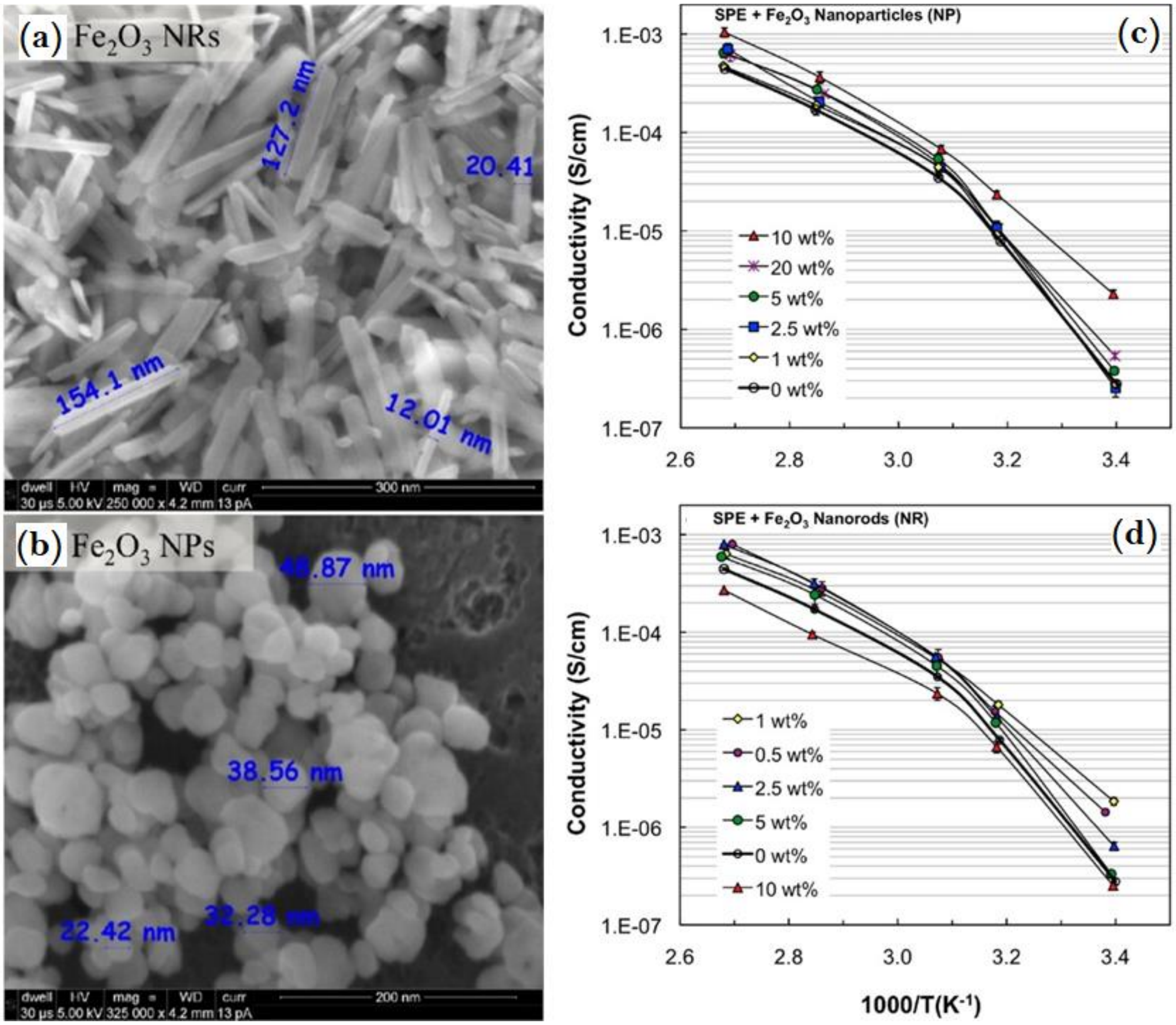

**Figure 21**. Scanning electron micrographs of (a) α-$Fe_2O_3$ nanorods (NR) with average length 105 ± 32 nm and average diameter 16 ± 5.4 nm and (b) α-$Fe_2O_3$ nanoparticles (NP) with average diameter 29 ± 11 nm. The error represents one standard deviation from the mean and Ionic conductivity versus temperature for PEO/$LiClO_4$ SPEs filled with (c) α-$Fe_2O_3$ NPs and (d) α-$Fe_2O_3$ NRs. The symbols represent the average of two measurements, and error represents the largest and smallest measured values. With permission from Ref. [79] Copyright © 2012, American Chemical Society.

Another effective approach is to create the positively charged oxygen vacancies on the 1 D nanofiller surface which acts as Lewis acid sites in the composite polymer electrolyte matrix. Liu et al., [80] reported the $Y_2O_3$-doped $ZrO_2$ (YSZ) nanowire (1 D NW; av. diameter 55 nm) addition in the PAN−$LiClO_4$ polymer electrolyte (Figure 22 a). The motto behind the use of nanowire was the formation of effective percolation network across a long distance. SEM micrograph of after the NW dispersion demonstrates the smooth morphology and evidence uniform distribution of nanowire in the host polymer matrix. The ionic conductivity was improved from 2.98×10-6 S $cm^{-1}$ (YSZ nanoparticle) to $1.07\times10^{-5}$ S $cm^{-1}$ (7 YSZ nanowire) (Figure 22 b). This increase was due to the formation of continuous pathways with nanowire dispersion having oxygen vacancies (Figure 22 c). This helps in migration of more cations participating in the conduction.  Further evidence was obtained from the FTIR deconvolution pattern of the anion ($ClO_4^-$) and it displays lowest ion pair for 7 YSZ and hence the highest ionic conductivity. This was in correlation with the increased $Li^+$ ion transference number from 0.27 to 0.56. From TGA analysis a correlation of ionic conductivity was built with dehydration temperature and lower the dehydration temperature higher will be the conductivity and lower crystallinity. The voltage stability window was increased as compared to the polymer salt matrix.

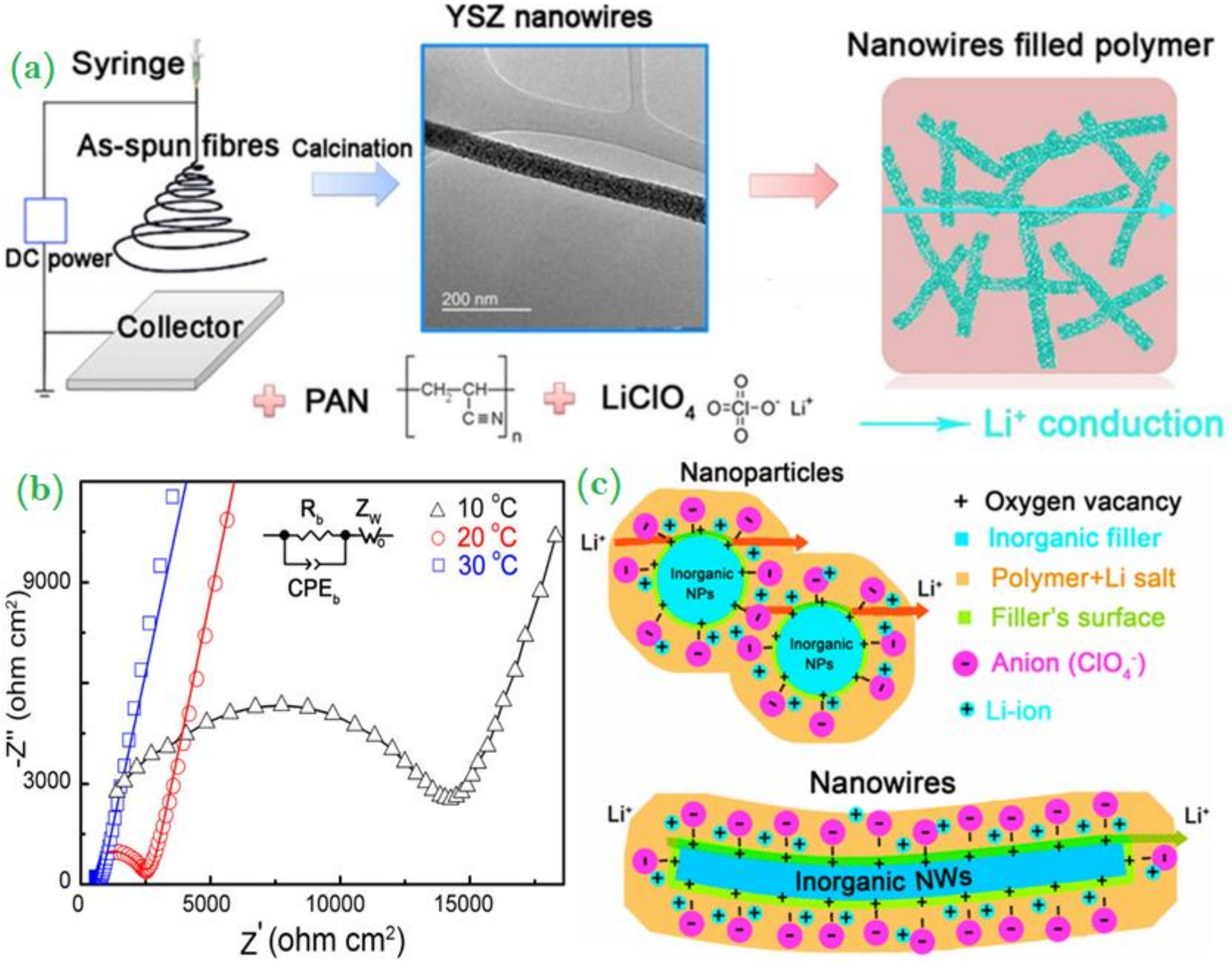

Figure 22. Schematic illustration for the synthesis of the solid composite polymer electrolyte. Electrospinning set up for the preparation of the YSZ nanowires, together with a TEM image of calcined nanowires. PAN, $LiClO_4$ and YSZ nanowires constitute the composite polymer electrolyte. Electrical properties of the composite polymer electrolyte

filled with the YSZ nanowires. (b) Experimental and fitting impedance spectra for the composite electrolyte with YSZ nanowires at different measuring temperatures and equivalent circuit. (c) Schematic illustration for Li-ion transport in the composite polymer electrolytes with nanoparticle and nanowire fillers. The positive-charged oxygen vacancies on the surfaces of the fillers act as Lewis acid sites that can interact strongly with anions and release Li-ions. A continuous fast conduction pathway can be seen for nanowires rather than nanoparticles. With permission from Ref. [80] Copyright © 2016, American Chemical Society.

Although the ionic conductivity was enhanced after the dispersion of the nanowire compared to nanoparticle but, one possibility still arise there that may affect the ion migration or restrict the effective role of nanowire/nanorod in the polymer matrix. As in the previous reports, the random distribution of the nanowire is investigated, but here if the nanowire is aligned parallel to the electrodes then they may show the negative effect by blocking the perpendicular ion migration. So, the alignment of the nanorod/nanowire may play and elective role in enhancing the conductivity further (Figure 23 a-d).

So, a recent report by Liu wt al., [81] investigated the effect of alignment of the nanowire (LLTO ; $Li_{0.33}La_{0.557}TiO_3$) in the three orientation ($0^o \pm 5^o$, $45^o \pm 9^o$ and $90^o \pm 8^o$) w.r.t of electrodes in a PAN-$LiClO_4$ based polymer matrix. SEM analysis depicts that the NW are well embedded in the polymer matrix with an average spacing of 5 μm (Figure 23 e-g). The ionic conductivity of randomly aligned NW was $5.40\times10^{-6}$ S $cm^{-1}$ and is higher than dispersed with the nanoparticle. Further when the NW were aligned with $0^o$ orientation then conductivity was increased to $6.05\times10^{-5}$ S $cm^{-1}$.

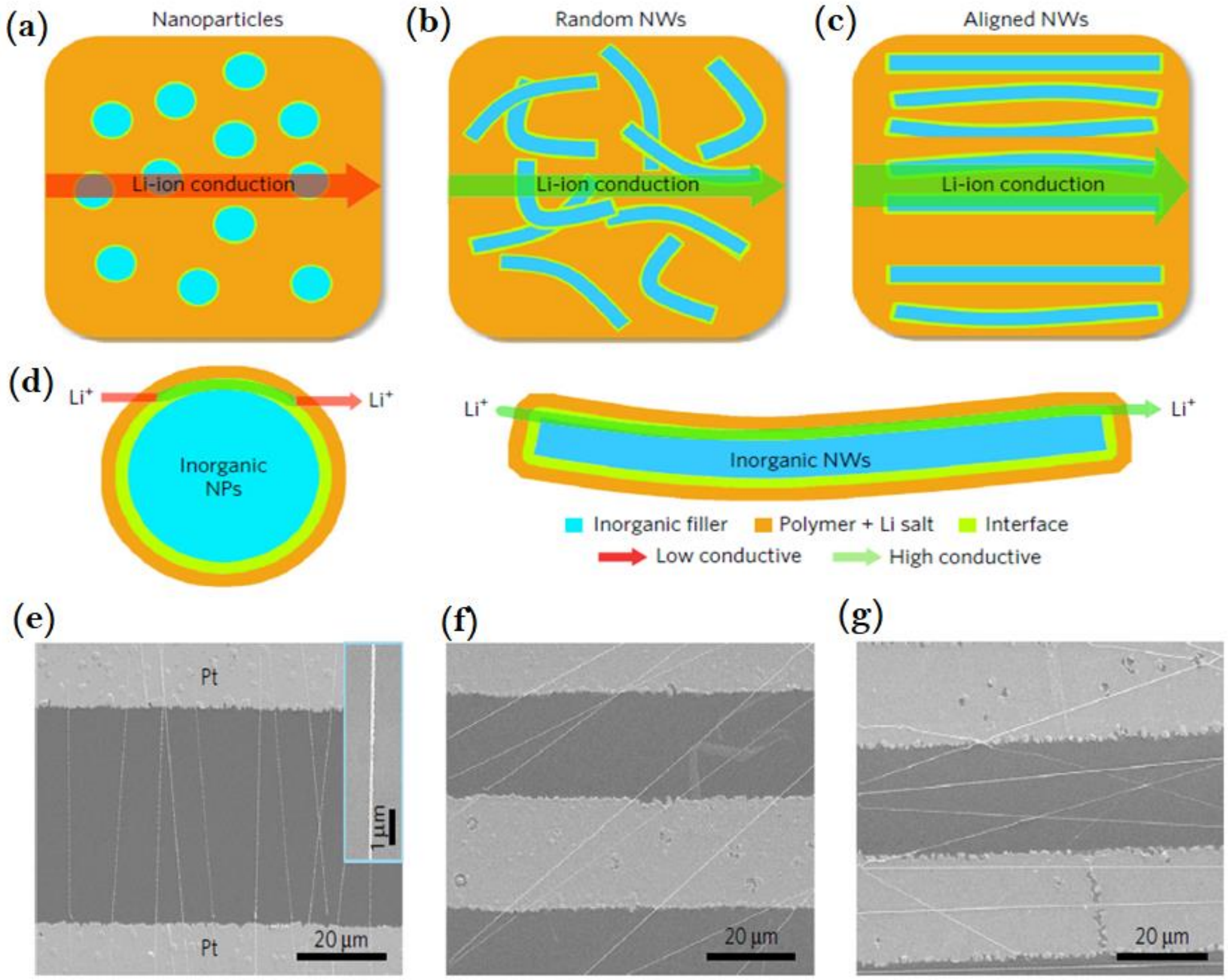

Figure 23. The comparison of possible Li-ion conduction pathways. a–c, Li-ion conduction pathways in composite polymer electrolytes with nanoparticles (a), random nanowires (b) and aligned nanowires (c). Compared with isolated nanoparticles, random nanowires could supply a more continuous fast conduction pathway for Li ion. Compared with random nanowires, aligned nanowires are free of crossing junctions. d, The surface region of inorganic nanoparticles (NPs) and nanowires (NWs) acts as an expressway for Li-ion conduction. e–g, SEM images of the aligned nanowires at orientations of $0^{o}$ (e), $45^{o}$ (f) and $90^{o}$ (g). The inset in Figure d is a SEM image at high magnification for the aligned nanowires. With permission from Ref. [81] Copyright © 2017 Springer Nature.

This enhancement in the conductivity may be due to the absence of the crossing junctions as in the nanoparticle and aligned NW. For orientation angle $45^{o}$ conductivity decreases to $2.24\times10^{-5}$ S $cm^{-1}$ and is due to more continuous length along electrodes as compared to the $0^{o}$. While for the $90^{o}$ orientation the conductivity was $1.78\times10^{-7}$ S $cm^{-1}$ and is due to the parallel alignment of NW with electrodes. Also, the electronic conductivity was of the order of $10^{-11}$ S $cm^{-1}$ and is negligible as compared to the ionic conductivity. The dispersion of the nanowire may also increase the polymer chain segmental motion and that was evidenced by the increased $Li^{+}$ transference number from 0.27 to 0.42. Further, the faster relaxation was confirmed by plotting the imaginary part of impedance and peak shifts toward high-frequency evidence the same. To further verify the experimental conductivity data simulation using the Comsol Multiphysics' (Comsol) numerical analysis of the current distribution was performed and is in good agreement with experimental data.

Another report by the Zhang et al., [82] demonstrates the synthetization of the composite gel polymer electrolytes by adding $SiO_2$ nanowires into a P(VDF-HFP) matrix. FESEM analysis concludes that the synthesized nanowire are flexible and of high purity with the diameter and length of the $SiO_2$ as <50 nm and ∼1 µm, respectively. FTIR was performed to check the nanorod formation and it shows the main absorption bands asymmetric bending mode of –Si–O–Si, symmetric bending mode of –Si–O–Si, bending mode of –O–Si–O and the bending mode of the –Si–OH. FESEM analysis of the composite matrix suggests the effective role in modifying the surface morphology of pure polymer which are favorable for cation transport. Also, the presence of pores suggests the absorption of more electrolyte and hence the more ionic conductivity. FTIR spectrum confirms the interaction between the Si atoms of $SiO_2$ and F atoms of P(VDF-HFP) chain which helps in ion migration. TGA plot reveals the improved thermal stability (427 $^{o}$C) as compared to the pure polymer (384 $^{o}$C) and is due to the interaction between the host polymer matrix and the Si nanowire. Further DSC analysis shows the reduction in the crystallinity on the addition of nanowire and may be due to the lowering of polymer chain reorganization tendency and enhanced amorphous content. The stress-strain curve shows increased tensile strength from 18.3 MPa (zero % nanowire) to 27.3 MPa for composite polymer electrolyte with 10 wt. % nanowire and is owing to the interaction between the polymer matrix and $SiO_2$ nanowire. This may be due to the increased chain flexibility on addition of nanowire which makes it capable of bear high stress and is in agreement with the FTIR. The ionic conductivity was increased by one order and is $1.08\times10^{-3}$ S $cm^{-1}$ (at 30 $^{o}$C) and may be due to the weakening of the interaction of cation with the fluorine of host polymer matrix. Also, the porous structure as evidenced in FESEM provides some additional conducting pathways in same volume favorable for ion transport and space charge region build up occurs due to the more free charge carriers availability. At, high content the interpenetrating networks are formed due to the overlapped space charge regions and hence the ion conducting pathways. Also, the transference number of $Li^{+}$ was increased from 0.23 to 0.70 on the addition of nanowire and may be due to the increased number of free ions via the dual interaction of acidic sites with fluorine of polymer host as well as an anion ($PF_6^-$). The electrochemical stability window of the synthesized composite polymer electrolyte was 4.8 V and no effect of a surface group of the nanofiller was observed on the stability window.

Another report based on $Mg_2B_2O_5$ nanowires (Av. diameter ~270 nm) with PEO-LiTFSI was reported by Sheng et al., [83]. The advantages with $Mg_2B_2O_5$ NWs are the hardness of 15.4 GPa and Young's modulus of 125.8 GPa [84]. The highest ionic conductivity was $1.53 \times 10^{-4}$ S cm$^{-1}$ (@40 °C) and $3.7 \times 10^{-4}$ S cm$^{-1}$ (@50 °C) for 10 wt. % $Mg_2B_2O_5$ nanowires. The enhancement was attributed to the coordination interaction of anion ($TFSI^-$) with $Mg_2B_2O_5$ nanowires and it enhances the salt dissociation which increases the number of lithium ions for migration (Figure 24 a). Another reason may be the reduction of the crystallinity and reorganization tendency after incorporation of NW which promotes faster segmental motion as evidenced by the XRD and DSC analysis. The lithium transference number ($t_{Li^+}$) was 0.44 and is much higher as compared to pure PEO which ($t_{Li^+}$ =0.19). The voltage stability window was 4.7 V and is superior to NW free system which shows 4.25 V.

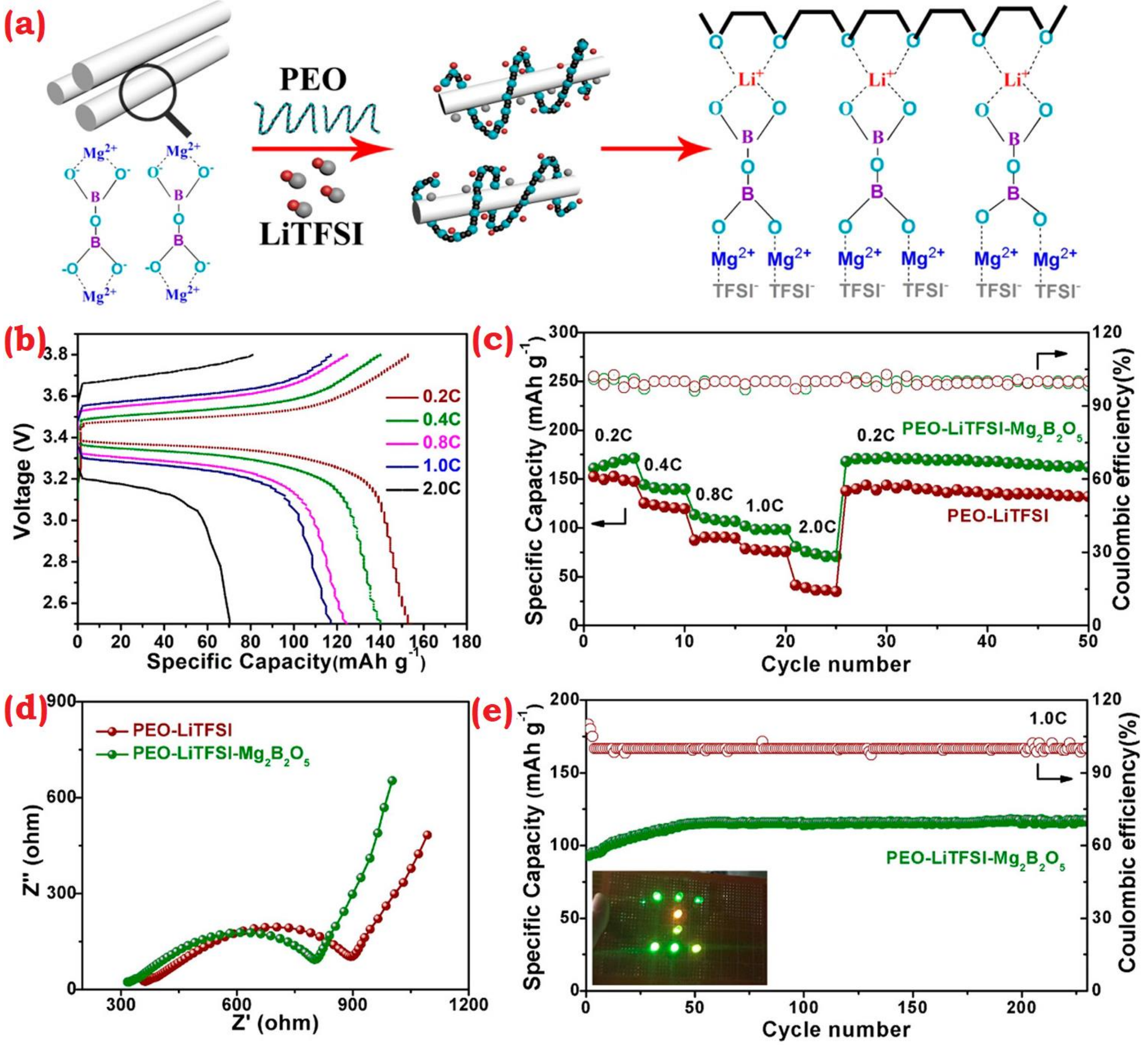

Figure 24. (a) Schematics of lithium ion migration in $Mg_2B_2O_5$ enhanced composite SSEs, (b) Typical charge−discharge curves of $LiFePO_4$/Li SSLIBs using PEO-LiTFSI-10 wt % $Mg_2B_2O_5$ electrolyte at 50 °C. (c) Rate performance of $LiFePO_4$/Li SSLIBs using SSEs with and without $Mg_2B_2O_5$ at 50 °C. (d) EIS spectra of battery using SSEs with and without $Mg_2B_2O_5$ at 50 °C. (e) Cycling performance of $LiFePO_4$/Li SSLIBs with PEO-LiTFSI-10 wt % $Mg_2B_2O_5$ electrolyte at 1.0 C and 50 °C. The inset is a digital photograph of LEDs lightened by SSLIBs. With permission from Ref. [83] Copyright © 2018 American Chemical Society.

The maximum strength was 2.29 MPa and mechanical properties were improved after addition of NW. The prepared system shows excellent flame-retardant performance. Figure 25 b-e shows the electrochemical performance of Li-ion battery assembled using PEO-LiTFSI- $Mg_2B_2O_5$. The discharge and charge voltage plateaus are around 3.35 and 3.50 V (@ 0.2 C) and over-potential between charge-discharge plateau increases with the increase of current density (Figure 24 b). Figure 24 c shows the performance with NW and without NW. The polymer electrolyte system with NW displays the specific discharge capacity of about 158 (@0.2 C), 140 (@0.4 c), 124 (@0.8 C), 117 (@1.0 C), and 72 mAh $g^{-1}$ (@2.0 C), while without NW the specific discharge capacity are 139 (@0.2 C), 121 (@0.4 c), 92 (@0.8 C), 75 (@1.0 C), and 37 mAh $g^{-1}$ (@2.0 C). It concluded that addition of NW increases the cyclic stability and capacity independent of current rate. Figure 24 d shows the impedance study of the battery with and without NW. Addition of NW reduces both charge-transfer resistance ($R_{ct}$) and the ohmic resistance ($R_o$) which indicates the increase of ionic conductivity with NW. Figure 24 e shows the stable specific capacity nearly 120 mAh $g^{-1}$ in 230 discharge−charge cycles (@ 1.0 C, 50 oC) with a Coulombic efficiency of 100%. The inset in the Figure 25 e shows a lightening LED with present battery (Shape of "I").

### 4.4. Separator Development: Commercial and Patents

As the electrolyte/separator is the key component of the battery. The first commercialized battery produced in 1991 by SONY Cooperation, comprised a porous plastic film soaked typically in $LiPF_6$ dissolved in a mixture of EC/EMC/DEC. The role of the separator is to provide path to ions, (i) cathode to the anode during charging, and (ii) anode to cathode during discharging. Table 4 summarizes the various manufacturers and composition of the separator and Table 5 (information available on their site) summarizes the properties of commercial separators.

**Table 4.** (a) Major Manufacturers of Lithium-Ion Battery Separators along with Their Typical Products. With permission from Ref. [85] Copyright © 2004, American Chemical Society.

| Manufacturer | Structure | Composition | Process | Trade name |
|---|---|---|---|---|
| Asahi Kasai | Single layer | PE | wet | HiPore |
| Celgard LLC | single layer | PP, PE | Dry | Celgard |
| | multilayer | PP/PE/PP | Dry | Celgard |
| | PVdF coated | PVdF, PP, PE, PP/PE/PP | Dry | Celgard |
| Entek Membranes | single layer | PE | Wet | Teklon |
| Mitsui Chemical | single layer | PE | Wet | - |
| Nitto Denko | single layer | PE | Wet | |
| DSM | single layer | PE | Wet | Solupur |
| Tonen | single layer | PE | Wet | Setela |
| Ube Industries | multi-layer | PP/PE/PP | dry | U-pore |
| Targary | - | PE/PP | wet | NAATBat |
| Daramic | - | PE | - | |
| Dry Lyte | - | Solid Polymer Electrolyte | | SEEO |
| Nanomyte®SE-50 | Self-standing (Amorphous) | Polymer-Ceramic Composite Material with Lithium Salt | In acetonitrile | NEI Co. |
| NANOMYTE® H-polymer | Self-standing (Amorphous) | PEO-based polymer electrolyte with lithium salt | In acetonitrile | NEI Co. |
| Ionic Materials | | Polymer Electrolyte | | Ionic mat. Inc. |

**Table 5.** Typical Properties of Some Commercial Microporous Membranes

| Property → Company ↓ | Thickness (average) | Basis Weight (g/m2) | Porosit y | Air Permeability | Tensile Strength | Tensile Elongation at Break | Puncture Resistance | Thermal Shrinkage @ 120 C, 1 hr. |
|---|---|---|---|---|---|---|---|---|
| ENTEK | 20 µm | 10.5 | 43 % | 300 sec | MD > 1,000 kg/cm2<br>TD - 750 kg/cm2 | MD - 90%<br>TD - 300% | 550 g | MD - <10%<br>TD - <3% |
| ENTEK EPX | 12 µm | 5.5 | 54% | 65 sec | MD > 1,000 kg/cm2<br>TD - 800 kg/cm2 | MD - 90%<br>TD - 230% | 300 g | MD - <12%<br>TD - <7% |
| EPENTEK EPH | 16 µm | 8.4 | 47% | 150 sec | MD > 1,000 kg/cm2<br>TD - 700 kg/cm2 | MD - 80%<br>TD - 250% | 430 g | MD - <10%<br>TD - <4% |

The separator must provide smooth ion migration and should be as thin as possible with lightweight, and must prevent electronic conduction. The most crucial point is that separator must have good compatibility, less thickness (< 25 μm) with the electrodes and possess desirable mechanical, electrochemical and stability (thermal/voltage) properties [85-86]. Table 6 summarizes the selected patents in polymer electrolytes.

**Table 6.** Some available patents on electrolytes.

| Inventor | Polymer Used | Patent No. | Year | Conductivity |
|---|---|---|---|---|
| Bauer et al | $LiClO_4$ in a 400 MW PEG | 4,654,279 | 1987 | $4\times10^{-4}$ S/cm at 25° C |
| Kuzhikalail M et al | PAN, EC, PC and $LiPF_6$/$LiAsF_6$ | 5510209 | 1995 | OCV:2.85 V |
| Nitash Pervez Balsara et al | Block copolymer | US 8,889,301 B2 | 2014 | $1\times10^{-4}$ S cm at 25°C |
| Wunder et al | PEO- POSS-phenyl7$(BF_3Li)_3$ | 9680182 B2 | 2017 | $1\times10^{-4}$ S cm at 25°C. (for O/Li=14) |
| Michael A. Zimmerman | PPS, PPO, PEEK, PPA | 2017/0018781 A1 | 2017 | $1\times10^{-5}$ S/cm (At RT) |
| Russell Clayton Pratt et al | perfluoropolyether electrolytes terminated with urethane | 9923245 | 2018 | $3.6 \times 10^{-5}$ (@40 ºC)<br>$1.1 \times 10^{-4}$ (@80 ºC) |
| Mohit Singh et al | Ceramic electrolyte | 20110281173 | 2018 | Stability upto 500 cycles |

**Summary and Conclusions**

This chapter highlighted the recent updates on the solid PNCs with a nanoparticle of different shape and specific surface area. The development of solid polymer electrolyte as an alternative to liquid/gel polymer electrolyte has been focused by researchers all over the world. From last three decades the nanoparticles have gained attention for preparation of the solid polymer electrolyte for the energy storage systems. Amongst the various nanoparticles used for dispersion in the polymer matrix, nanoclay, active/passive nanofiller have been most studied. The nanofiller and nanoclay improved the overall properties of the polymer electrolytes. The nanofiller enhanced the electrical properties along with mechanical properties depending on the preparation method, dispersion, surface group interaction with polymer chains, surface area and dielectric constant. The Lewis aid base interaction mechanism was elaborated for the understanding the cation transport. The serious drawback with nano filler was that at high concentration it is not able to play its effective role in enhancing the properties may be due to the possibility of aggregation. Also, the lack of a continuous path for cation limits the enhancement in the rate of ion transport. In order to resolve the issue of agglomeration search of new nanoparticle ended with the nanoclay. The advantage with the nanoclay is that percolation threshold is lower as compared to nano filler and it fit for the elimination of most detrimental factor i.e. concentration polarization. It may consider as the first step toward the realization of single ion solid polymer electrolyte (PNC). As the specific surface area plays an effective role in enhancing the electrical and transport properties along with mechanical properties. So, the research was focused toward the use of nanoparticle with the high specific surface area and sufficient oxygen vacancies that promotes the faster ion transport. This lead the development of nanorod, nanowire as the dispersive element in the solid polymer electrolytes. One remarkable advantage with the nanorod was that a long continues path was available for ion migration. This provides the smoother ion migration between the electrodes along with improved mechanical and thermal properties. The nanowire dispersion is the recent advancement in solid polymer electrolytes. The oxygen vacancies on the nanowire surface provide additional sites to the ion for a long time. This enhances the ionic conductivity and thermal/mechanical properties. But, one barrier was the random alignment of the nanowire that hinders the ion migration. This was solved by the alignment of the nanowire and it provides continues path to the ion between the electrodes without any constraint. The electrochemical cells fabricated using the nanorod, nanowire promises their launch at a commercial level.

**Acknowledgments:** The author thanks the Central University of Punjab for providing fellowship.